\documentclass[aip,pop,reprint,superscriptaddress,longbibliography,showkeys]{revtex4-2}
\usepackage{lmodern}
\usepackage[T1]{fontenc}
\usepackage[utf8]{inputenc}
\usepackage{mathrsfs}
\usepackage{amsmath}
\usepackage{amsthm}
\usepackage{amssymb}
\usepackage{graphicx}
\usepackage{latexsym}
\usepackage{bm}
\usepackage{amsfonts}
\usepackage{babel}
\usepackage{enumitem}
\usepackage[normalem]{ulem}
\usepackage{soul}
\usepackage{subcaption}
\usepackage{makecell}

\usepackage{tikz}
\usetikzlibrary{shapes.geometric, arrows}

\tikzstyle{io} = [trapezium, 
trapezium stretches=true, 
trapezium left angle=80, 
trapezium right angle=100, 
minimum width=3cm, 
minimum height=1cm, 
text centered, 
text width=4.1cm, 
draw=black
]

\tikzstyle{process} = [rectangle, 
minimum width=3cm, 
minimum height=1cm, 
text centered, 
text width=3.2cm, 
draw=black]

\tikzstyle{decision} = [diamond, 
minimum width=3cm, 
minimum height=1cm, 
text centered, 
aspect=2,
draw=black]

\tikzstyle{arrow} = [thick,->,>=stealth]

\begin{document}

\title{Statistical inference of anomalous thermal transport with uncertainty quantification for interpretive 2-D SOL models}

\author{Y. Fu}
\email{fu9@llnl.gov}
\author{B. D. Dudson}
\author{X. Chen}
\author{M. V. Umansky}
\author{F. Scotti}
\author{T. D. Rognlien}
\affiliation{Lawrence Livermore National Laboratory, Livermore, CA 94550, USA}

\author{A. W. Leonard}
\affiliation{General Atomics, PO Box 85608, San Diego, CA 92186, USA}

\date{\today}

\begin{abstract}
The critical task of inferring anomalous cross-field transport coefficients is addressed in simulations of boundary plasmas with fluid models. A workflow for parameter inference in the UEDGE fluid code is developed using Bayesian optimization with parallelized sampling and integrated uncertainty quantification. In this workflow, transport coefficients are inferred by maximizing their posterior probability distribution, which is generally multidimensional and non-Gaussian. Uncertainty quantification is integrated throughout the optimization within the Bayesian framework that combines diagnostic uncertainties and model limitations. As a concrete example, we infer the anomalous electron thermal diffusivity $\chi_\perp$ from an interpretive 2-D model describing electron heat transport in the conduction-limited region with radiative power loss. The workflow is first benchmarked against synthetic data and then tested on H-, L-, and I-mode discharges to match their midplane temperature and divertor heat flux profiles. We demonstrate that the workflow efficiently infers diffusivity and its associated uncertainty, generating 2-D profiles that match 1-D measurements. Future efforts will focus on incorporating more complicated fluid models and analyzing transport coefficients inferred from a large database of experimental results. 
\end{abstract}


\maketitle

\section{Introduction}

In modern magnetic fusion devices, like tokamaks, the magnetic field lines are closed within the separatrix and open outside. The open field line region between the separatrix and the solid armor material of the chamber wall is called the scrape-off layer (SOL) and is commonly referred to as the boundary of magnetic fusion devices \cite{stangeby2000plasma,krasheninnikov2020edge}. 2-D transport models, such as SOLPS \cite{wiesen2015new} and UEDGE \cite{rognlien1999two}, are commonly used to study SOL physics, including predicting plasma detachment \cite{coster2011detachment,jaervinen2018b} and estimating heat flux width \cite{ballinger2021simulation,emdee2021predictive}. In 2-D models, electrons, ions, and impurities are modeled as fluids based on the Braginskii equations \cite{braginskii1965transport}, and neutrals are modeled as another fluid species or kinetic particles. Although plasma transport parallel to the magnetic field is well captured in the Braginskii equations, perpendicular transport is usually anomalous and thought to be primarily due to plasma turbulence \cite{krasheninnikov2008recent}. In 2-D models, anomalous transport is usually represented by energy and particle diffusion coefficients, which are user-defined inputs for the model. These transport coefficients are typically tuned to match experimental measurements, a process known as ``interpretive modeling.'' However, the coefficient-tuning process can be tedious, time-consuming, and may be subject to human bias. Therefore, an automated transport coefficient estimation procedure is becoming increasingly important for efficient and accurate modeling of SOL plasma.

There has been a long effort to automate the estimation of the transport coefficient in 2-D SOL transport models. The objective of coefficient estimation is to minimize the discrepancy between the model prediction and experimental measurements, which is usually quantified by a loss function and minimized using iterative methods. In early development of SOLPS, an algorithm was implemented to minimize the loss function defined as the mean square error in temperature, density profiles, and their gradients \cite{coster2000automatic,schneider2006plasma}. A gradient descent method was implemented to search for the minimum, where the gradient was calculated using finite differences. This method was later improved \cite{carli2022bayesian} by employing a posterior distribution based on the Bayesian framework as a loss function and using automatic differentiation. Fixed-point iteration methods were also developed \cite{canik2011measurements} to tune transport coefficients iteratively, avoiding the need to calculate the gradients in each step. Other iterative methods were also explored, such as the autoUEDGE \cite{izacard2018automatic,nelson2021interpretative} code that adjusts the transport coefficients to match the measured spatial gradients of temperature and density. Iterative methods are useful in practice, but have some disadvantages: They are vulnerable to local minima that may differ from the global optimum value that represents the true solution. These methods may also fail to identify multiple solutions to the system because the transport coefficients may not be uniquely defined. In addition, when transport coefficients are adjusted, the underlying 2-D transport models may no longer converge, which could prevent iterative methods from proceeding further. Iterative methods also don't provide a measure of uncertainty in the result.

Recently, a normalizing flow model based on neural networks was built \cite{furia2022normalizing} to map out the probability distribution of transport coefficients within a given range, which could overcome some of the difficulties of iterative methods. However, mapping out the full distribution function might be too expensive when the focus is solely on a few optimum transport coefficients. In this paper, a new method for estimating transport coefficients using Bayesian optimization (BO) \cite{2018arXiv180702811F, garnett_bayesoptbook_2023} with the Bayesian framework \cite{von2011bayesian,kruger2024thinking} is presented. BO is a global optimization method for expensive-to-evaluate black-box functions. In recent years, it has been widely applied to many optimization and design tasks in fusion research, such as studying core transport \cite{rodriguez2018vitals,rodriguez2022nonlinear,rodriguez2024enhancing}, neutral density profile \cite{bogar2021direct}, rampdown scenario \cite{mehta2024automated}, disruption mitigation \cite{pusztai2023bayesian,ekmark2024fluid}, divertor component design \cite{humphrey2023machine}, numerical simulation calibration \cite{jarvinen2022bayesian,crovini2024automatic} in tokamaks, and for coil configuration optimization for stellarators \cite{glas2022global,giuliani2024direct}. Similarly, in inertial fusion research, BO has been used for laser and target optimization, as well as experimental design \cite{mariscal2024toward,li2023hybrid,wang2024multifidelity,gammel2024gaussian}.
The Bayesian framework allows us to construct the probability distribution of transport coefficients, which is generally a high-dimensional non-Gaussian distribution, based on prior knowledge and a likelihood function. This work utilizes BO to find the optimum transport coefficient in UEDGE that maximizes the probability distribution. Uncertainties are also estimated based on the approximated probability distribution function obtained during the BO process. A batch-parallelized sampling method is used to improve computational efficiency.

Parallelized BO has several potential advantages over iterative transport-estimation methods. Though not completely avoidable, BO is less likely to be stuck in a local minimum due to it being based on global optimization. BO can also be used to identify multiple viable solutions to a problem. Additionally, since multiple 2-D transport models are run in parallel, the algorithm will continue to run even when some cases do not converge. It is also less expensive than traditional Bayesian inference \cite{von2011bayesian} since the Gaussian process (GP) approximates the distribution function, whose evaluation is much faster than iteratively running a fluid code like UEDGE in the SOL domain of a medium-to-large tokamak..

The remainder of the paper is organized as follows: Section~\ref{sec:BO} briefly discusses the basic strategy of BO and its parallelization. A comprehensive review of BO and related topics can be found in Ref.~\onlinecite{garnett_bayesoptbook_2023}. Section~\ref{sec:BO_transport_estimation} presents details of the transport coefficients estimation in UEDGE and the overall workflow. Section~\ref{sec:interpretive_model} introduces a simple interpretive model describing the electron heat transport dominated by conduction and radiation losses, where the cross-field heat diffusivity $\chi_\perp$ needs to be estimated. Section~\ref{sec:synthetic_validation} validates our method using synthetic data with known $\chi_\perp$ to demonstrate the self-consistency of the inference method. The inference process is applied to experimental data from discharges run on the DIII-D tokamak in Section~\ref{sec:experimental_data}. Conclusion and discussion are presented in Section~\ref{sec:conclusion}.

\section{Bayesian optimization
\label{sec:BO}}

BO is a large class of optimization methods that seeks the global minimum of an expensive-to-evaluate objective function $f$ in the parameter space $\Sigma \subset \mathbb{R}^N$, defined as:
\begin{equation}
    x^* = \underset{x\in \Sigma}{\mathrm{argmin}}\,f(x).
    \label{eq:optimization}
\end{equation}
In the task of transport coefficient estimation, $x\in\mathbb{R}^N$ is an array of $N$ transport coefficients, and $f(x)$ is the loss function describing the discrepancy between the profiles produced by the UEDGE code, such as radial electron temperature and density profiles, and experimental observations. BO searches the optimum solution by strategically evaluating the function $f(x_i)$ with $x_i\in \Sigma$, which can be summarized as follows: Assume that we initially sampled $n$ points $\{x_i\}$ in the parameter space and obtained corresponding results $\{f(x_i)\}$. BO provides the most promising point $x_{n+1}$ that $f(x_{n+1})$ will be the optimum of $f(x)$ based on the previous $n$ observed points. After running the UEDGE with transport coefficients $x_{n+1}$, a new data point $f(x_{n+1})$ is gathered. Then, BO can predict the next point $x_{n+2}$ to continue the process till some exit criteria are reached. 

BO consists of two main ingredients to find the most promising point at each step. Firstly, BO uses a cheap surrogate model, usually GP regression \cite{williams2006gaussian,michoski2024gaussian}, to approximate the expensive function $f(x)$, in this case solutions from UEDGE. After fitting existing observations $\{f(x_i)\}$, the GP can predict the most probable value and uncertainty of $f(\tilde{x})$ for any unobserved point $\tilde{x}\in \Sigma$, which are expressed as $\mu(\tilde{x})$ and $\sigma(\tilde{x})$, respectively. Next, BO utilizes both $\mu(\tilde{x})$ and $\sigma(\tilde{x})$ to construct an acquisition function $A(\tilde{x})$. The acquisition function balances the exploitation of observed data and the exploration of unobserved regions, which can be defined based on various criteria such as the upper confidence bound (UCB) \cite{srinivas2009gaussian}, the probability of improvement (PoI) \cite{jones2001taxonomy}, and the expected improvement (EI) \cite{jones1998efficient}. The argument of the maximum of the acquisition function, $\tilde{x}^* = \mathrm{argmax}\,A(\tilde{x})$, represents the most promising point to be evaluated in the next step. Since evaluating a fitted GP and its derivatives is fast, the search for the optimum in $A(\tilde{x})$ is very efficient.

Although the BO procedure seems sequential, it can be easily parallelized to increase search efficiency. After fitting the current observations with a surrogate model, more complicated acquisition functions can be used to predict multiple points to evaluate in the next step. One of the simplest methods is the constant liar approximation \cite{ginsbourger2010kriging}, which can be described as follows. After an acquisition function provides a single point $\tilde{x}^*$ based on $n$ existing observations $\{f(x_i)\}$, constant liar assumes that the value $f(\tilde{x}^*)$ has been observed and approximates it by a constant, which can be the mean, minimum, or maximum of previously observed data. Then, the acquisition function predicts the next point $\tilde{x}^{**}$ based on $n$ true and $1$ fictitious observations $\{f(x_i); \mu(\tilde{x}^*)\}$. Similarly, to generate the $m$-th observation point, we use $n$ true observations and $m-1$ fictitious observations. This procedure continues until the desired number of points is generated. More complicated parallelization strategies can also be applied, such as the multipoints expected improvement \cite{ginsbourger2010kriging} and the maxvar acquisition method \cite{jarvenpaa2019efficient}. These methods can also be evaluated asynchronously, where new samples start immediately after one or a few calculations finish.

\section{BO for transport coefficients estimation
\label{sec:BO_transport_estimation}}

UEDGE uses the implicit Newton-Krylov method to find the equilibrium in a set of nonlinear partial differential equations (PDE) for any given set of transport coefficients. When successfully converged, UEDGE provides 2-D profiles, such as temperature and density, of electrons, ions, and impurities. Loss functions are then calculated to quantify the discrepancy between experimental observations and profiles produced by UEDGE  at the same locations as measured by diagnostics in the fusion device. This process of calculating profiles and loss is known as the forward problem. Depending on the setting, such as the number of species and physical effects included, the forward problem can be very time-consuming. Finding the optimum transport coefficients that minimize the loss is the inverse problem, which can be addressed by the BO method. This section presents the basic workflow of using parallelized BO within the Bayesian framework to estimate transport coefficients in UEDGE. The implementation of BO is realized using the \texttt{scikit-learn} \cite{pedregosa2011scikit} and \texttt{BayesianOptimization} \cite{BayesianOptimization} libraries. The BO workflow for transport coefficients estimation has been incorporated into the UEDGE Toolbox, \texttt{UETOOLS}. The physical model used to validate this workflow will be presented in the next section.

\subsection{Bayesian framework}

This section discusses the Bayesian framework \cite{von2011bayesian,kruger2024thinking} and constructs the loss function. Let $\mathcal{D}$ be the experimental data with measurement uncertainty, such as the electron temperature measured at many points along a radial chord at the outer midplane (OMP) of a fusion device; $\mathcal{M}$ be the model we considered, such as a UEDGE case with a given magnetic geometry and boundary condition; $\theta$ be an array of parameters to be inferred. The probability of parameters $\theta$ with given data $\mathcal{D}$ and model $\mathcal{M}$ can be written as $P(\theta|\mathcal{D},\mathcal{M})$, also known as the posterior distribution. Based on Bayes' theorem, such a probability can be written as 
\begin{equation}
    P(\theta|\mathcal{D},\mathcal{M}) = \dfrac{P(\mathcal{D}|\theta,\mathcal{M}) P(\theta|\mathcal{M})}{P(\mathcal{D}|\mathcal{M})},
\end{equation}
where $P(\mathcal{D}|\theta,\mathcal{M})$ is the likelihood, $P(\theta|\mathcal{M})$ is the prior distribution, and $P(\mathcal{D}|\mathcal{M})$ is a normalization constant known as the evidence.

The likelihood $P(\mathcal{D}|\theta, \mathcal{M})$ indicates the probability of observing the data $\mathcal{D}$ given the parameter $\theta$ and the model $\mathcal{M}$. The specific form of the likelihood will be discussed later after we introduce the observation data. The prior distribution $P(\theta|\mathcal{M})$ represents prior knowledge of the parameters $\theta$ before inference, which is purely based on the researcher's experience or assumption. The transport coefficients in the SOL can sometimes vary spatially by several orders of magnitude \cite{nelson2021interpretative}, so we assume that the prior of the order of magnitude of $\theta$, i.e., $\ln\theta$, has a uniform distribution, implying that the prior of $\theta$ is log-uniform. Since the array of transport coefficients $\theta$ are positive, we can define $\vartheta \doteq \ln \theta$. Let $\vartheta\in[\vartheta_a, \vartheta_b]$, the prior can be written as
\begin{equation}
    P(\vartheta|\mathcal{M}) = \dfrac{1}{\vartheta_b - \vartheta_a}
    \quad\Rightarrow\quad
    P(\theta|\mathcal{M}) = \dfrac{1}{\vartheta_b - \vartheta_a} \dfrac{1}{\theta}. 
    \label{eq:log_uniform}
\end{equation}

The evidence $P(\mathcal{D}|\mathcal{M})$ does not depend on $\theta$, so it acts as a normalization constant. Although the evidence does not affect the optimization, the value $P(\mathcal{D}|\mathcal{M})$ itself is important in the task of model selection \cite{chilenski2019importance}. The higher the value of $P(\mathcal{D}|\mathcal{M})$, the better the model $\mathcal{M}$ describes the experimental data $\mathcal{D}$.

We define our loss function $\mathcal{L}$ as proportional to the negative logarithm of $P(\theta|\mathcal{D}, \mathcal{M})$ so that the optimal parameter $\theta$ maximizes the posterior distribution. This is known as the maximum a posteriori (MAP) estimate. For convenience, we work on the magnitude $\vartheta\doteq \ln\theta$ to take advantage of its constant prior. In practice, $\ln[P(\vartheta|\mathcal{D},\mathcal{M})]$ typically varies across several orders of magnitude, which causes some difficulty for optimization algorithms. Therefore, we define the loss function as
\begin{equation}
    \mathcal{L}(\vartheta) \doteq \ln \left\{ - \ln[P(\vartheta|\mathcal{D},\mathcal{M})]  \right\}.
    \label{eq:loss_function}
\end{equation}
The task of finding the optimum coefficients is equivalent to the minimization task 
\begin{align}
    \vartheta^* = \underset{\vartheta\in \Sigma}{\mathrm{argmin}} \,\mathcal{L}(\vartheta).
    \label{eq:optimization2}
\end{align}

\subsection{Uncertainties quantification}

The optimum coefficient $\vartheta^*$ is the only return variable when using traditional gradient-based methods to solve optimization problems like Eq.~(\ref{eq:optimization2}). In comparison, BO returns not only the optimum $\vartheta^*$ but also an approximate loss function $\mathcal{L}(\vartheta)$ using a surrogate model. Therefore, we can construct an approximate posterior distribution $\tilde{P}(\vartheta|\mathcal{D},\mathcal{M})$. Although the error bar from experimental data is usually assumed to be Gaussian, the posterior distribution $P(\vartheta|\mathcal{D},\mathcal{M})$ is generally non-Gaussian due to the nonlinearity of the UEDGE model and the complexity of the experimental data. Nevertheless, we can roughly quantify the uncertainty of the estimated coefficients $\vartheta^*$ as
\begin{equation}
    \xi \doteq \left( \int (\vartheta - \vartheta^*)^2 \tilde{P}(\vartheta|\mathcal{D},\mathcal{M}) \mathrm{d}\vartheta \right)^{1/2},
    \label{eq:uncertainty}
\end{equation}
which can be understood as the standard deviation of a random variable.

\begin{figure}[b]
    \centering
    \includegraphics[width=0.7\columnwidth]{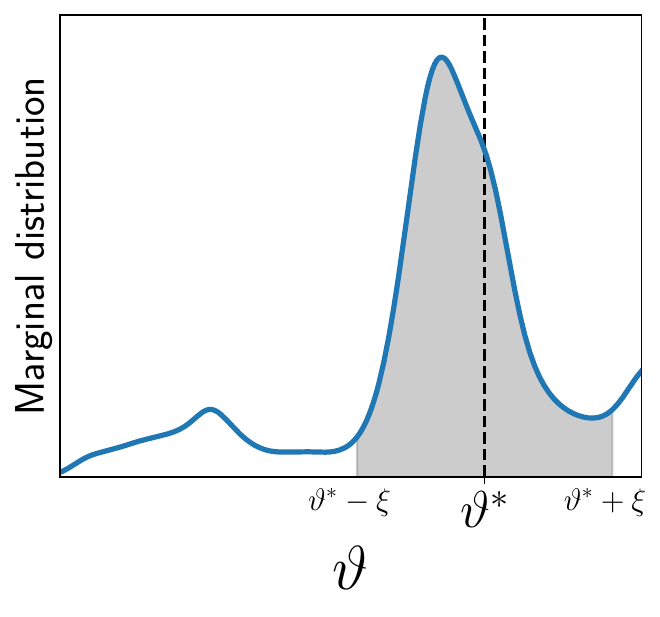}
    \caption{Demonstration of a 1-D marginal distribution. The maximum of the full distribution is indicated by $\vartheta^*$, which is different from the maxima of the marginal distribution. The uncertainty calculated by Eq.~(\ref{eq:uncertainty}) is indicated by $\vartheta^*\pm\xi$.}
    \label{fig:distribution_demo}
\end{figure}

Heuristically, the narrower the posterior distribution $P(\vartheta|\mathcal{D}, \mathcal{M})$ is around its maximum $\vartheta^*$, the less uncertain the estimation of the coefficient is, and vice versa. Multiple local maxima of $P(\vartheta|\mathcal{D}, \mathcal{M})$ will also contribute to the formula. Assume that the global maxima is $\vartheta^*$ while another local maxima is at $\vartheta^{**}$. If ${P}(\vartheta^*|\mathcal{D}, \mathcal{M})$ and ${P}(\vartheta^{**}|\mathcal{D}, \mathcal{M})$ have similar probabilities, the presence of local maxima will lead to a large uncertainty. Otherwise, if ${P}(\vartheta^*|\mathcal{D}, \mathcal{M}) \gg {P}(\vartheta^{**}|\mathcal{D}, \mathcal{M})$, the local maxima have almost no contribution to the uncertainty. In addition, the maxima of a high-dimensional distribution may differ from the maxima of its 1-D marginal distribution, which also increases the uncertainty in Eq.~(\ref{eq:uncertainty}). A specific case, obtained in the inference task in Section~\ref{sec:inference_results}, is shown in Fig.~\ref{fig:distribution_demo} as a demonstration. Therefore, the magnitude of $\xi$ depends on how concentrated $P(\vartheta|\mathcal{D}, \mathcal{M})$ is around its global maxima and whether there are multiple local maxima with similar probability.

The primary objective of BO is to find the global optimum so that the surrogate model may be more accurate in high-probability regions and less accurate elsewhere. The approximated posterior distribution $\tilde{P}(\vartheta|\mathcal{D}, \mathcal{M})$ obtained from BO may not be as accurately estimated as other probability density estimation methods, such as in Ref.~\onlinecite{furia2022normalizing}. However, an accurate approximation of the high-probability regions is sufficient when calculating the transport coefficients and their uncertainty. Therefore, BO is advantageous due to its faster evaluation speed.

\subsection{Convergence of UEDGE
\label{sec:convergence}}

One notable difficulty in estimating the transport coefficient is the convergence of UEDGE when evaluating the loss function $\mathcal{L}(\vartheta)$. For a given parameter $\vartheta\doteq \ln\theta$, the UEDGE case may fail to converge for various reasons. For example, when the initial state of variables is too far from the equilibrium, the Newton-Krylov method may not find a solution of nonlinear PDEs. Furthermore, if the model $\mathcal{M}$ and the parameters $\vartheta$ are not physically compatible, such as when an imbalance exists between input and output power, equilibrium can never be found. 

Fortunately, the parallelized BO method has some tolerance for the convergence failures. In each step, when many points $\{\vartheta_j\}$ are sampled and only a small portion of the samples fail to converge, we can update the surrogate model only with the converged samples and continue the BO procedure. In addition, for those unconverged cases, one can use the continuation solver in \texttt{UETOOLS} to get a ``partial solution'' described as follows. Consider the case where we want to find a equilibrium with a new parameter $\tilde\vartheta$, but UEDGE cannot converge directly. We can search for the existing converged equilibrium in previous iteration to find a parameter $\vartheta_0$ closest to $\tilde\vartheta$. The continuation solver starts from the equilibrium with $\vartheta_0$ and calculates a new equilibrium with gradually increasing parameter $\vartheta=\vartheta_0 + \lambda (\tilde\vartheta - \vartheta_0)$ with $\lambda\in[0,1]$. When increasing $\lambda$ from 0 to 1, UEDGE may no longer converge at some point with $\lambda^*$. Then, we can use the last convergent equilibrium with parameter $\vartheta^*=\vartheta_0 + \lambda^* (\tilde\vartheta - \vartheta_0)$ to calculate the loss function $\mathcal{L}(\vartheta^*)$. Although it is not an evaluation of the loss function at $\tilde\vartheta$ provided by the acquisition function, we can still use it as an alternative and update the surrogate model.

Nevertheless, when UEDGE models $\mathcal{M}$ are too complicated, parallelized BO may still have difficulty due to convergence issues. In particular, if most of the sample points $\{\vartheta_j\}$ do not converge and the continuation solver only provides $\lambda^* \ll 1$, the BO procedure cannot continue. In those cases, one needs to seek other methods to estimate transport coefficients. For example, one can use a multi-fidelity approach by first limiting the number of inference coefficients $\vartheta$, finding proper transport coefficients, and then returning to the more complicated model with previously inferred coefficients fixed. 

\begin{figure}[t]
    \centering
    \begin{tikzpicture}[node distance=1.6cm]
        \node (in1) [io] {Define UEDGE model $\mathcal{M}$ and inference coefficients $\vartheta$};
        \node (in2) [io, below of=in1, yshift=0.2cm] {Choose observations $\mathcal{D}$ and define likelihood $P(\mathcal{D}|\vartheta, \mathcal{M})$};
        \node (pro1) [process, below of=in2, yshift=0.1cm] {Run UEDGE with $n$ initial samples and get losses $\{\mathcal{L}(\vartheta_i)\}$};
    
        \node (pro2) [process, below of=pro1] {Use all data $\{\mathcal{L}(\vartheta_i)\}$ to build a surrogate model};
        \node (dec1) [decision, below of=pro2, yshift=-1.5cm] {Exit criteria};
        \node (pro3) [process, right of=dec1, xshift=2.75cm, text width=3cm] {Calculate next points $\{\tilde{\vartheta}_j\}$ from acquisition function};
        \node (pro4) [process, above of=pro3, text width=3cm] {Run UEDGE in parallel to get data $\{\mathcal{L}(\tilde{\vartheta}_j)\}$};
        \node (pro5) [process, right of=pro2, xshift=2.75cm, text width=3cm] {Combine $\{\mathcal{L}(\tilde{\vartheta}_j)\}$ with existing data};
    
        \node (out1) [io, below of=dec1, text width=3.5cm] {Return the optimum $\vartheta^*$ and its uncertainty $\xi$};
        
        \draw [arrow] (in1) -- (in2);
        \draw [arrow] (in2) -- (pro1);
        \draw [arrow] (pro1) -- (pro2);
        \draw [arrow] (pro2) -- (dec1);
        \draw [arrow] (dec1) -- node[anchor=south] {no} (pro3);
        \draw [arrow] (pro3) -- (pro4);
        \draw [arrow] (pro4) -- (pro5);
        \draw [arrow] (pro5) -- (pro2);
        \draw [arrow] (dec1) -- node[anchor=east] {yes} (out1);
    \end{tikzpicture}
    \caption{The workflow of transport coefficient estimation with Bayesian optimization.}
    \label{fig:BO_workflow}
\end{figure}
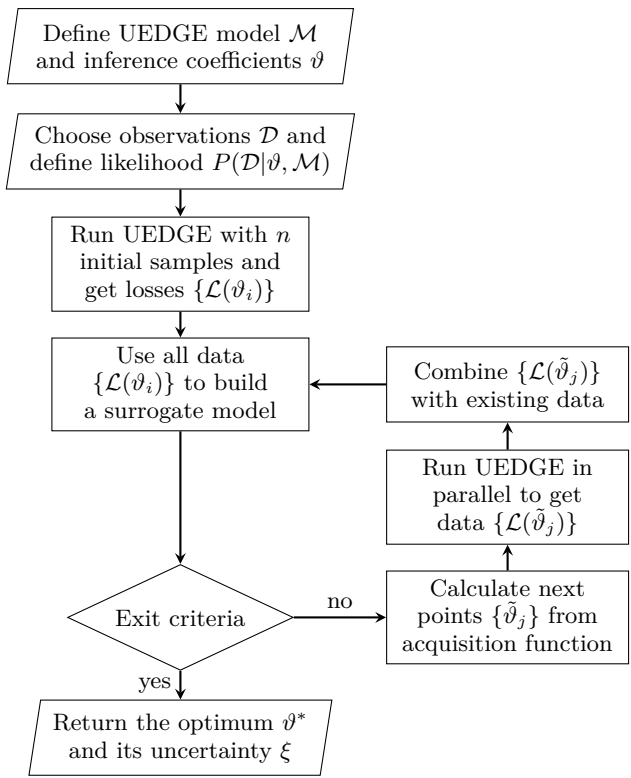

\subsection{Transport coefficients estimation workflow}

Here, we summarize the workflow for transport coefficient estimation in UEDGE, depicted in Fig.~\ref{fig:BO_workflow}. Firstly, we define the UEDGE model $\mathcal{M}$ to be used. It includes physical settings, such as particle species, physical effects, magnetic geometry, and boundary conditions, and numerical settings, such as grid size and numerical schemes. The number of transport coefficients $\vartheta$ that need to be inferred is also determined here. Then, we choose what experimental observation $\mathcal{D}$ to fit, such as the temperature and density profiles at the OMP. The likelihood function $P(\mathcal{D}|\vartheta, \mathcal{M})$ is then constructed using these experimental data to calculate the loss function $\mathcal{L}(\vartheta)$. After running UEDGE and calculating losses with some initial samples, the BO procedure begins to approximate the loss function by surrogate models and search for its global optimum. Finally, when some exit criteria are met, the optimum coefficients $\vartheta^*$ are returned with their uncertainty $\xi$ calculated from the approximate posterior distribution.

\section{Interpretive SOL models for electron heat transport
\label{sec:interpretive_model}}

The estimation of transport coefficients in complicated UEDGE models may encounter convergence issues, discussed in the previous section, and can be resource and time-intensive. We adopt a simple interpretive model of electron heat transport to test and benchmark our BO workflow. The interpretive model, which is a 2-D extension of the widely used 2-point model \cite{stangeby2000plasma}, will be fitted with experimental observations for a series of discharges on the DIII-D tokamak \cite{fenstermacher2022diii}. 

\begin{figure}[t]
    \centering
    \includegraphics[width=0.7\columnwidth]{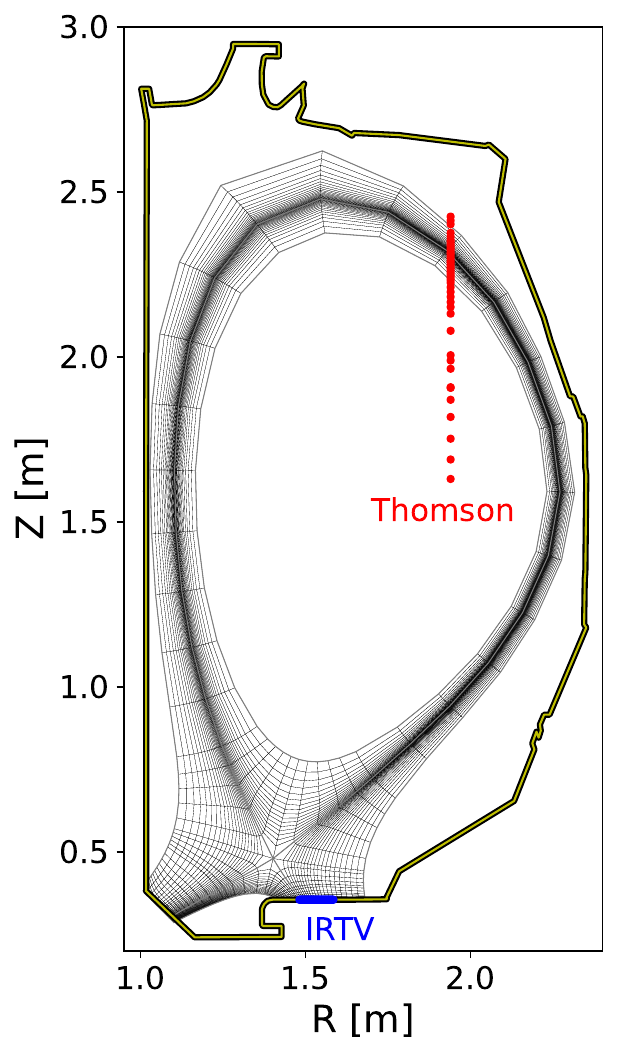}
    \caption{Overview of UEDGE grids based on DIII-D discharge \# 189051 at 3250 ms. The locations of Thomson scattering and infrared television (IRTV) camera measurements are shown in red and blue dots, respectively.}
    \label{fig:grid_and_diagnostic}
\end{figure} 

Consider the electron heat transport in the conduction-limited region dominated by conduction and radiation loss in the steady state:
\begin{equation}
    \nabla_\parallel q_\parallel + \nabla_\perp q_\perp = P_\mathrm{rad}, 
    \label{eq:electron_heat_transport}
\end{equation}
where $P_\mathrm{rad}$ is the radiated power density. $q_\parallel$ is the flux-limited parallel electron heat flux \cite{stangeby2000plasma}:
\begin{equation}
    \dfrac{1}{q_\parallel} = \dfrac{1}{q_\mathrm{Sp}} + \dfrac{1}{q_\mathrm{fl}}.
    \label{eq:flux_limiter}
\end{equation}
The Spitzer-dominated heat flux is
\begin{equation}
    q_\mathrm{Sp}\doteq \kappa_\parallel \nabla_\parallel T_\mathrm{e},
\end{equation}
where $T_\mathrm{e}$ is the electron temperature and $\kappa_\parallel\sim T_\mathrm{e}^{5/2}$ is the Spitzer thermal conductivity. The flux-limited heat flux describes the free-streaming heat flux as 
\begin{equation}
    q_\mathrm{fl} \doteq \alpha_\mathrm{e} n_\mathrm{e} \sqrt{e T_\mathrm{e} / m_\mathrm{e}}\, e T_\mathrm{e}, 
    \label{eq:limited_heat_flux}
\end{equation}
where $n_\mathrm{e}, m_\mathrm{e}$ is electron density and mass, $e$ is the elementary charge, and $\alpha_\mathrm{e}=0.21$ is an empirical constant. $q_\perp$ is the perpendicular electron heat flux:
\begin{equation}
    q_\perp \doteq n_\mathrm{e} \chi_\perp \nabla_\perp T_\mathrm{e}, 
\end{equation}
where $\chi_\perp$ is the anomalous perpendicular thermal diffusivity that needs to be inferred by our BO procedure. The convection heat transport in the parallel direction is ignored, while convection in the perpendicular direction is effectively included in the perpendicular diffusivity \cite{umansky1999empirical}. A typical UEDGE grid is shown in Fig.~\ref{fig:grid_and_diagnostic} with $N_x = 64$ mesh grids in the poloidal direction and $N_y=36$ in the radial direction. The boundary conditions set a fixed power $P_\mathrm{core}$ transporting across core boundary, set a fixed temperature at the target plate and private flux (PF) wall as $2\,\mathrm{eV}$ \footnote{Empirically, our result is not sensitive to the target plate temperature, as long as it is low enough}, and set the fixed temperature at the outer wall based on measured data.

The radiated power density, $P_\mathrm{rad}(R,Z)$, is obtained from 2-D bolometer tomography \cite{leonard19952d}, which is interpolated to UEDGE grids and serves as a fixed energy sink in Eq.~(\ref{eq:electron_heat_transport}). Using a fixed radiation profile has several advantages. Firstly, it captures one of the major energy sinks in electron heat transport, which makes the model more realistic. More importantly, using a fixed radiation profile is more computationally efficient than other models since equations for other impurity species are not needed. The fixed-fraction impurity model is computationally efficient in calculating the radiated power \cite{nelson2021interpretative}, but is also known to produce bifurcations of cold and hot branches \cite{krasheninnikov1995thermal,wising1996simulation,wigram2018uedge}. Therefore, the fixed-fraction model is not suitable for the black-box optimization method we used here. 

The momentum balance in the parallel direction determines the electron density. In the conduction-limited region, we have \cite{stangeby2000plasma}:
\begin{equation}
    \nabla_\parallel ( T_\mathrm{e} n_\mathrm{e} ) = 0.
    \label{eq:parallel_momentum}
\end{equation}
We set the radial electron density profile at the OMP equal to the Thomson scattering measurement \cite{ponce2010thomson, eldon2012initial} and then calculate its 2-D profile throughout the SOL using Eqs.~(\ref{eq:electron_heat_transport}) and (\ref{eq:parallel_momentum}). The electron density in the private flux region is set to $1\times 10^{19}\,\mathrm{m}^{-3}$ as it is not important for our inference. Due to the usage of the parallel momentum balance, no density boundary condition is needed.

Eq.~(\ref{eq:electron_heat_transport})-(\ref{eq:parallel_momentum}) are relatively simple since there are only two variables, $T_\mathrm{e}$ and $n_\mathrm{e}$, that need to be solved. However, due to the geometry and the nonlinearity in $\kappa_\parallel$ with respect to $T_\mathrm{e}$, analytical solutions are not available. Nevertheless, the model captures the most important parts of electron heat transport and can be used to efficiently infer the unknown perpendicular thermal diffusivity. 

\begin{figure}[t]
    \centering
    \includegraphics[width=\columnwidth]{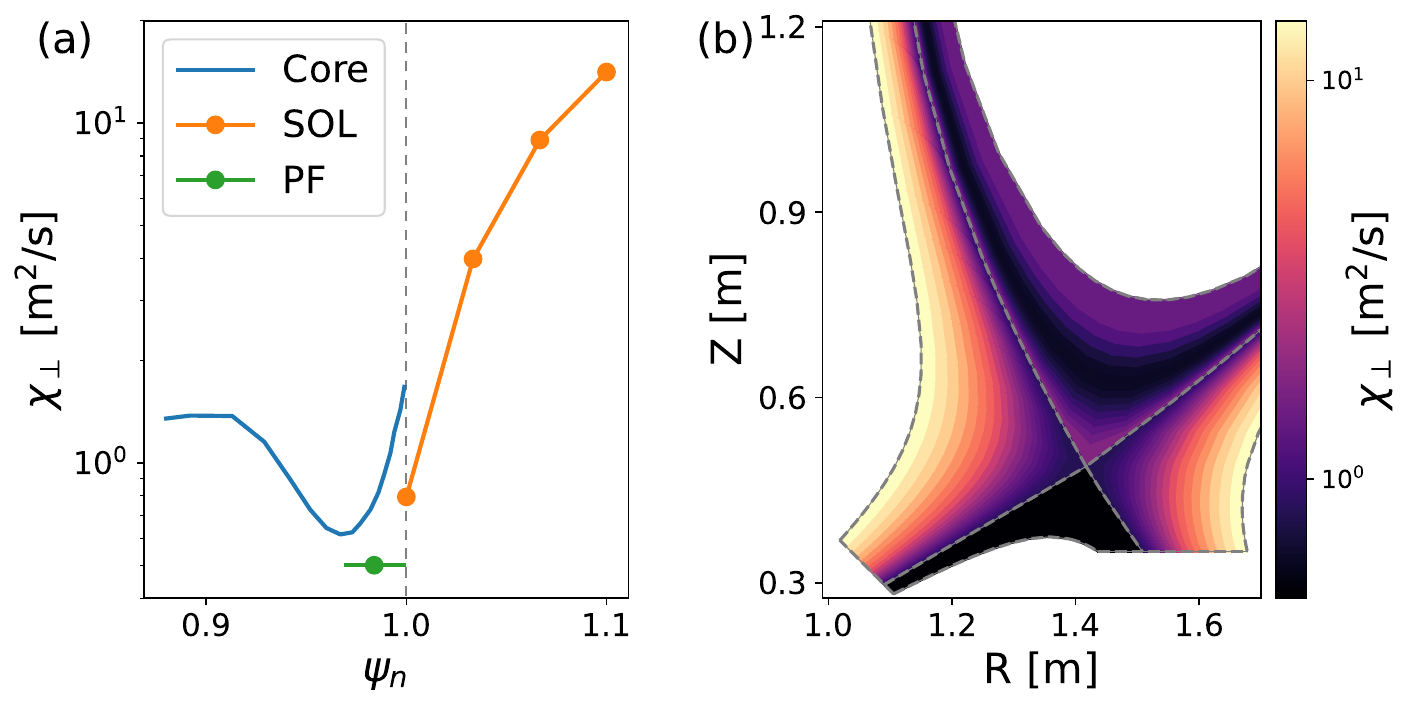}
    \caption{Demonstration of the $\chi_\perp$ profiles in (a) 1-D and (b) 2-D domain. The continuous profile in the core (closed field line) region is calculated interpretively using Eq.~(\ref{eq:interpretive_chi}). The piecewise linear profiles in SOL and private flux are represented by four and one constants, respectively.}
    \label{fig:chi_profile_demo}
\end{figure}

To simplify inference processes, we make a few assumptions about the $\chi_\perp$ profile. Although $\chi_\perp$ varies both in the radial and in the poloidal directions, there are not enough experimental measurements to detect its poloidal variation. So, $\chi_\perp$ is assumed to be a flux function with no variation in the poloidal direction. In the core region inside the separatrix, we can calculate $\chi_\perp$ interpretively \cite{callen2010analysis} from experimental data. Let $\psi$ be the poloidal magnetic flux function satisfying $|\nabla \psi| = B_\mathrm{p}R$, where $B_\mathrm{p}$ is the poloidal magnetic field. The interpretive $\chi_\perp$ is given by:
\begin{equation}
\chi_\perp(\psi) = 
- \dfrac{P_\mathrm{core} / A }{n_\mathrm{e} \langle B_\mathrm{p}R \rangle (\mathrm{d} T_\mathrm{e} / \mathrm{d} \psi)_\mathrm{fit}}.
\label{eq:interpretive_chi}
\end{equation}
Here, $A(\psi)$ is the area of $\psi$ flux surface, and $\langle\cdot\rangle$ is the flux surface average, e.g., the average of $B_\mathrm{p} R$ for the entire poloidal extent. $\mathrm{d} T_\mathrm{e} / \mathrm{d} \psi$ is obtained by fitting the experimental data using a GP regression and calculating its derivative from finite difference \cite{gammel2024gaussian}. In the SOL and private flux regions, we represent the $\chi_\perp$ profile by four and one constants, respectively, as a function of radial location with a simple linear interpolation in between, which is demonstrated in Fig.~\ref{fig:chi_profile_demo}. Here, $\psi_n$ indicates the normalized poloidal magnetic flux, with $\psi_n=1$ representing the separatrix. The number of parameters can be increased at the cost of a longer optimization time. The task of selecting the best number of parameters is known as model selection \cite{chilenski2019importance}, which is beyond the scope of the present article and will be left for future work. Therefore, the unknown parameters are reduced to a set of discrete values to be inferred by our BO workflow.

\section{Validation on synthetic data
\label{sec:synthetic_validation}}

To ensure the consistency of our inference method, we benchmark it against synthetic data generated from a given heat diffusivity $\chi_0$ as ground truth. We show that our inference process can successfully recover the ground truth within the estimated uncertainty. Other interesting details of the BO process, such as the trajectory of the inferred optimum parameter and its multi-dimensional probability distribution function, will also be discussed.

\subsection{Synthetic experiment setup
\label{sec:synthetic_setup}}

\begin{figure}[t]
    \centering
    \includegraphics[width=0.9\columnwidth]{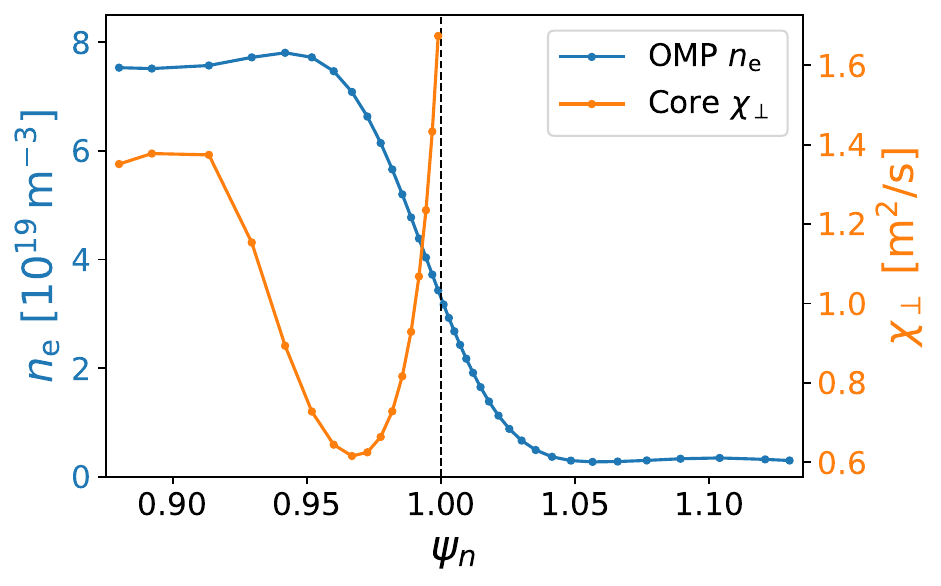}
    \caption{Synthetic electron density $n_\mathrm{e}$ and heat diffusivity $\chi_\perp$ profiles. $n_\mathrm{e}$ span the whole OMP, while $\chi_\perp$ only covers plasma in the closed field line region.}
    \label{fig:synthetic_ne_chi}
\end{figure} 

This section presents our setup following the workflow in Fig.~\ref{fig:BO_workflow}. In the benchmark case, the plasma shape is chosen to be a lower single-null configuration based on DIII-D discharge \# 189051, shown in Fig.~\ref{fig:grid_and_diagnostic}. The core power is chosen to be $6~\mathrm{MW}$, a typical value for DIII-D discharges. For simplicity, we turned off the radiated power loss during the benchmark. Figure~\ref{fig:synthetic_ne_chi} shows our synthetic OMP density profile and core $\chi_{\perp}$ profiles, which are chosen to be H-mode-like with an edge transport barrier, characterized by a narrow region of low $\chi_\perp$ around $0.9<\psi_n<1$. Those two profiles are fixed during the inference process. As the ground truth, we chose, somewhat arbitrarily, the heat diffusivity in the private flux as $\chi_\mathrm{pf} = 10~\mathrm{m^2/s}$ and in SOL as 
\begin{equation}
    \begin{split}    
        \chi_\mathrm{sol} & = (\chi_\mathrm{sol,1}, \,\chi_\mathrm{sol,2}, \,\chi_\mathrm{sol,3}, \,\chi_\mathrm{sol,4}) \\[2pt]
        & = (10^{0.5}, 10^{1.25}, 10^{1.75}, 10^2)~\mathrm{m^2/s}.
    \end{split}
\end{equation}
at $\psi_n = (1., 1.04, 1.08, 1.12)$. Therefore, the variables to be inferred are $\theta=(\chi_\mathrm{pf},\chi_\mathrm{sol})$, which is assumed to be in a 5-D parameter space $\theta\in[10^{-2}, 10^3]^5\subset \mathbb{R}^5$. To utilize our log-uniform prior, inference is performed after a log transform as $\vartheta=\ln\theta$. 

\begin{figure}[t]
    \centering
    \includegraphics[width=0.7\columnwidth]{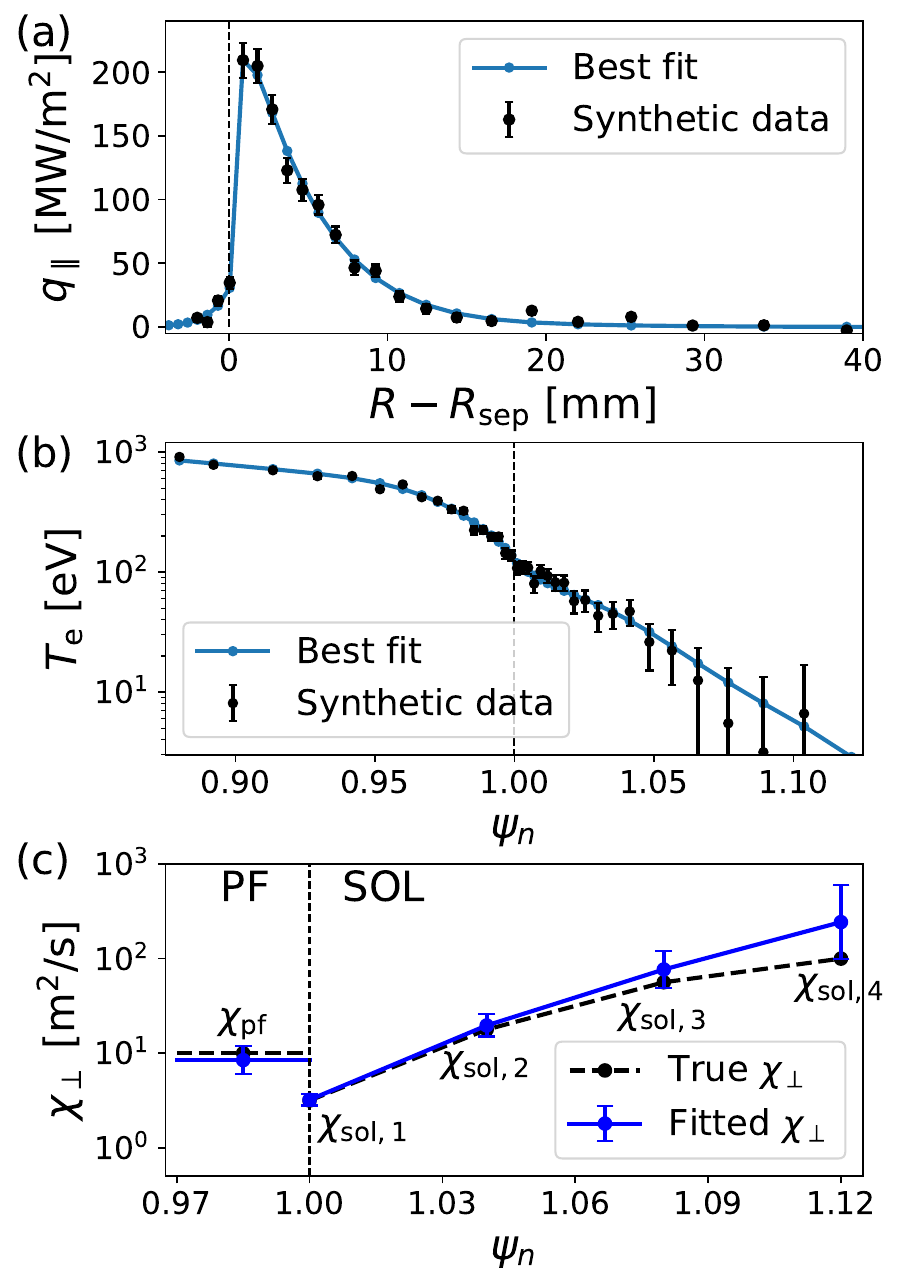}
    \caption{Benchmark result against synthetic data. (a) parallel heat flux in the outer divertor plate, which is magnetically mapped to the OMP. $R_\mathrm{sep}$ is the radius of the separatrix. (b) OMP electron temperature profile, (c) heat diffusivity profile. The dashed lines indicate the location of the separatrix.}
    \label{fig:synthetic_result}
\end{figure}

After UEDGE finds a convergent solution, we record the electron temperature profile $T_\mathrm{e}$ at the OMP and the parallel heat flux profile $q_\parallel$ at the outer divertor plate, the values of which are denoted $\{\mu_i\}$. We add Gaussian noise to the profiles with error bars $\{\sigma_i\}$ to represent error in the measured values, which is defined as 
\begin{equation}
    \sigma_i = 0.05 \mu_i + \delta.
    \label{eq:synthetic_error_bar}
\end{equation}
$\delta$ is a small constant (10~eV and 3~$\mathrm{MW/m^2}$, chosen empirically) to emulate uncertainties that do not scale with the measured value, so that data points with small measured values have relatively larger error bars. Our synthetic data combine the profile and its error as $\mathcal{D}=\{\mu_i, \sigma_i\}$, which are shown in Fig.~\ref{fig:synthetic_result} When given a parameter $\theta$, let $\{y_i\}$ be the profiles predicted by our physics model $\mathcal{M}$. We can define the $\chi^2$-metric
\begin{equation}
    \chi^2 \doteq \sum_{i=1}^N \dfrac{(y_{i} - \mu_{i})^2}{\sigma_{i}^2},
    \label{eq:chi_square}
\end{equation}
which indicates how accurately data $\mathcal{D}$ are matched by the new profile $\{y_i\}$. The likelihood function can be written as a normal distribution:
\begin{equation}
    P(\mathcal{D}|\vartheta,\mathcal{M}) = \exp\left(-\dfrac{\chi^2}{2}\right) \prod_{i} (\sqrt{2\pi} \sigma_{i})^{-1},
    \label{eq:likelihood}
\end{equation}
which consists of the loss function $\mathcal{L}(\vartheta)$ using Eq.~(\ref{eq:loss_function}).

Finally, some details of the BO process are presented as follows. With the loss function defined, we randomly sample 64 points $\{\vartheta_i\}$ in the 5-D space using the quasi-Monte Carlo method with the Sobol sequence \cite{caflisch1998monte} to calculate the loss $\{\mathcal{L}(\vartheta_i)\}$. Despite its relatively large sample size, all samples can be calculated independently in parallel with a relatively short wall time. Batch-parallelized BO is then performed. In each step, surrogate models based on GP regression with the Matern-5/2 kernel \cite{genton2001classes} are used to fit all available samples $\{\mathcal{L}(\vartheta_i)\}$. Two acquisition functions, EI and UCB, are used to predict 4 parameters $\{\tilde{\vartheta}_i\}$ in parallel using the constant liar approximation, respectively. If some parameters are too close to each other ($|\tilde{\vartheta}_i-\tilde{\vartheta}_j|<0.2$), duplicated points are abandoned, and new parameters are generated by tuning the hyperparameters of acquisition functions till 8 parameters are found. The losses $\{\mathcal{L}(\tilde{\vartheta_i})\}$ at these 8 parameters are calculated in parallel, which usually end within a similar amount of time ($\sim4$ min) due to the relative simplicity of the physics model. The next step will begin after all 8 calculations are finished, and the optimization process is terminated after 30 steps. In each step, we obtain an optimal parameter $\vartheta^*$ and a GP regression model of the posterior distribution $\tilde{P}(\vartheta|\mathcal{D},\mathcal{M})$. The uncertainty in each step $\xi$ can be calculated using the quasi-Monte Carlo integration of Eq.~(\ref{eq:uncertainty}). The whole inference takes around $2\sim 3$ hours to complete in total.

\subsection{Benchmark result}

\begin{figure}[b]
    \centering
    \includegraphics[width=\columnwidth]{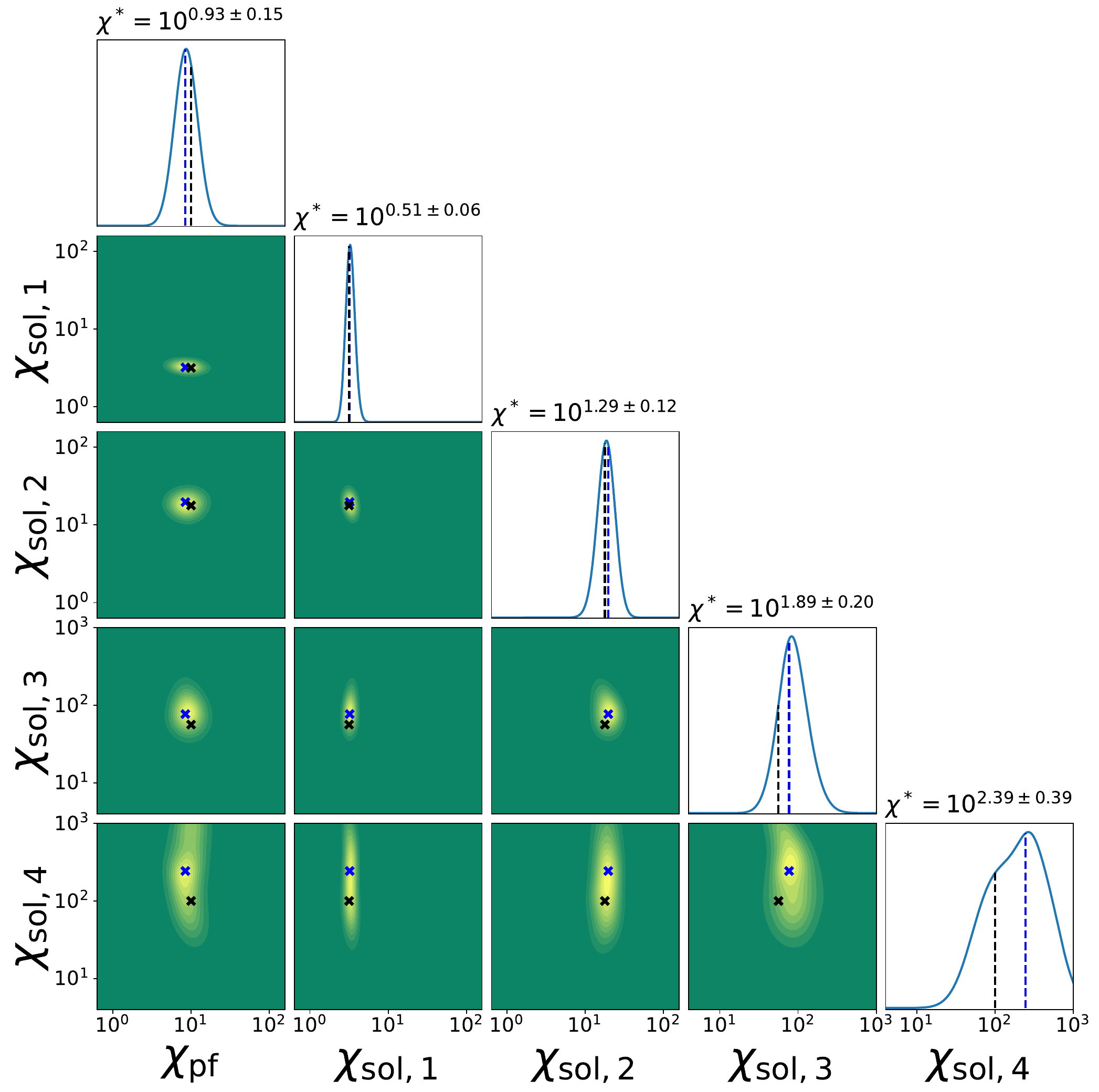}
    \caption{The pair plot of five inference parameters $\chi_\mathrm{pf}, \chi_\mathrm{sol,1}, ..., \chi_\mathrm{sol,4}$ in the unit $[\mathrm{m^2/s}]$. Diagonal (from top left to lower right) plots show 1-D marginal distribution and off-diagonal plots show joint 2-D marginal distributions of each parameter pair. The inferred parameters are shown in blue dashed curves in 1-D plots and blue crosses in 2-D plots, while the true parameters are shown in black dashed curves and black crosses.}
    \label{fig:pair_plot}
\end{figure} 

First, we present the inference result and compare it with the ground truth. Figure~\ref{fig:synthetic_result}(a) shows the parallel heat flux in the outer divertor plate, where the fitted profile coincides with the synthetic heat flux in terms of both the peak and the width of the heat flux. The fitted electron temperature profile at the OMP, shown in Fig.~\ref{fig:synthetic_result}(b), also overlays the synthetic profile on both sides of the separatrix. The inferred heat diffusivity is shown in Fig.~\ref{fig:synthetic_result}(c), which generally reproduces the true value within its uncertainties. However, the heat diffusivity in five locations is estimated to have different uncertainties as a result of their different degrees of influence on the profiles. Because the peak of the heat flux is located just outside the separatrix, its shape is most strongly affected by the $\chi_\perp$ near the separatrix. In addition, the temperature profile in the far SOL has a relatively larger error bar in Eq.~(\ref{eq:synthetic_error_bar}), which also increases the uncertainty of $\chi_\perp$ in that region. Therefore, $\chi_\perp$ just outside the separatrix, $\chi_\mathrm{sol,1}$, is the best fit with the smallest uncertainty, while $\chi_\perp$ in the far SOL, $\chi_\mathrm{sol,4}$, is the worst fit with the largest uncertainty. 

The uncertainty in Fig.~\ref{fig:synthetic_result}(c) is a simple characteristic of the 5-D posterior distribution $\tilde{P}(\vartheta|\mathcal{D},\mathcal{M})$. A pair plot is shown in Fig.~\ref{fig:pair_plot} to present more details of the distribution function. The diagonal figures show five 1-D marginal distributions of each parameter. The off-diagonal figures show ten joint 2-D marginal distributions of each pair of parameters. It shows that the distribution of $\chi_\mathrm{sol,1}$ is very narrow, while the distribution of $\chi_\mathrm{sol,4}$ is widespread, which is consistent with the uncertainty in Fig.~\ref{fig:synthetic_result}. In addition, unlike other parameters, the 1-D distribution of $\chi_\mathrm{sol,4}$ is non-Gaussian, with a relatively high probability density that extends from $10^{1.5} \sim 10^{2.8}$ (around $30 \sim 630$). This means that $\chi_\mathrm{sol,4}$ has a minimal effect on the profiles and, therefore, can only be estimated with a large uncertainty. 

\begin{figure}[b]
    \centering
    \includegraphics[width=0.9\columnwidth]{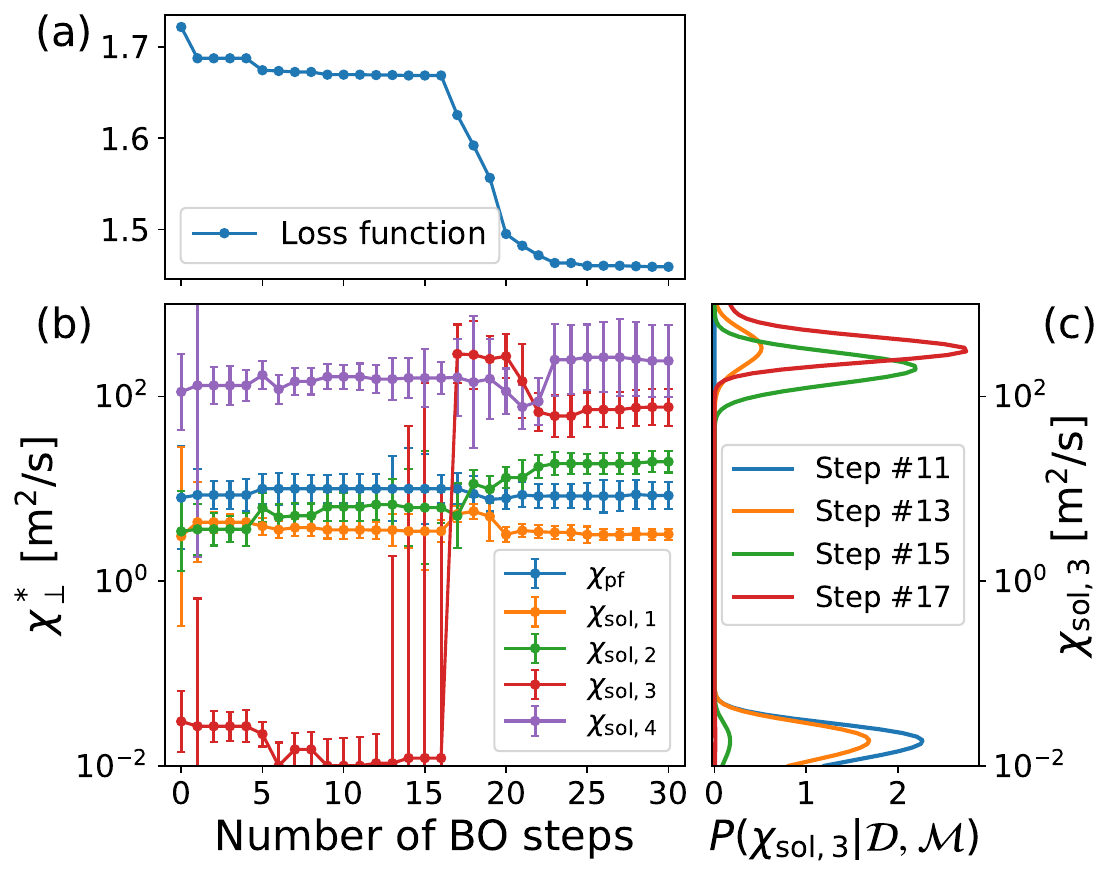}
    \caption{The trajectory in each step during the inference process: (a) loss function, (b) optimal parameters and their uncertainty, (c) marginal distribution of $\chi_\mathrm{sol,3}$ from steps 11 to 17.}
    \label{fig:bo_history}
\end{figure}

\begin{figure*}
    \centering
    \begin{subfigure}{.3\textwidth}
        \centering
        \includegraphics[width=\textwidth]{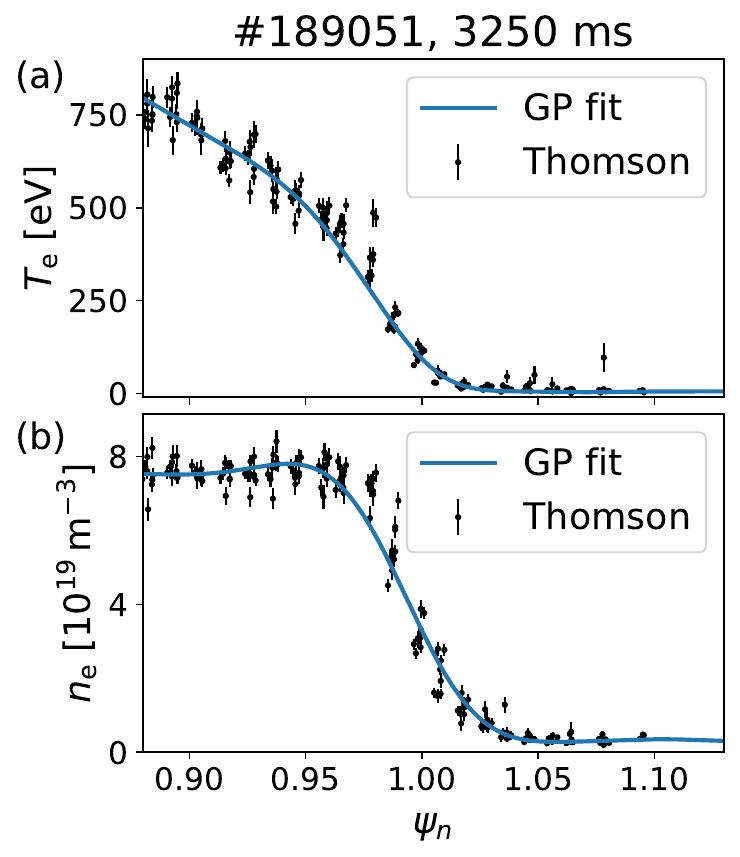}
    \end{subfigure}
    \hfill
    \begin{subfigure}{.295\textwidth}
        \centering
        \includegraphics[width=\textwidth]{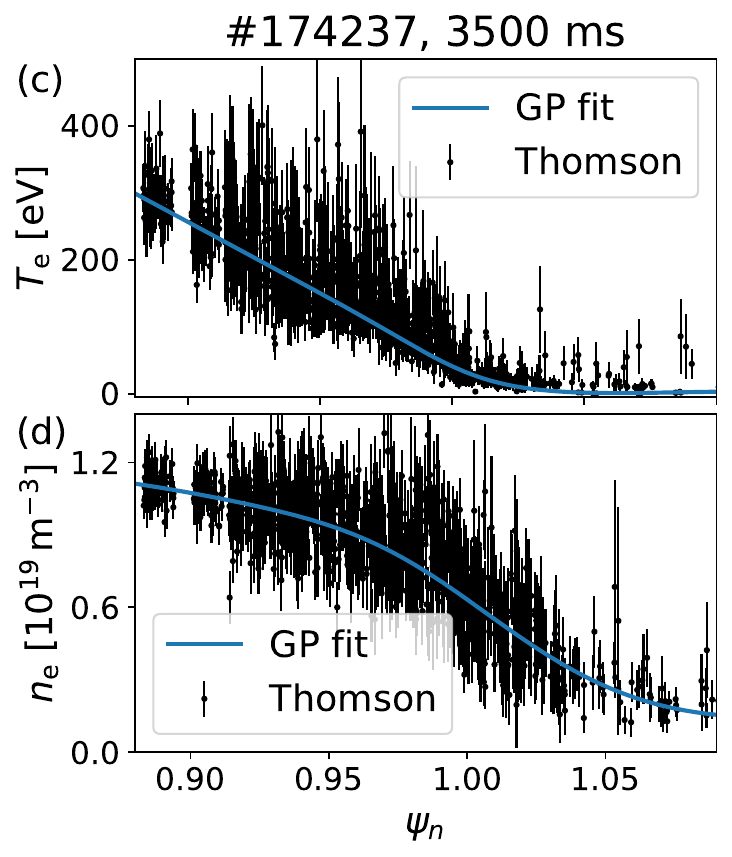}
    \end{subfigure}
    \hfill
    \begin{subfigure}{.3\textwidth}
        \centering
        \includegraphics[width=\textwidth]{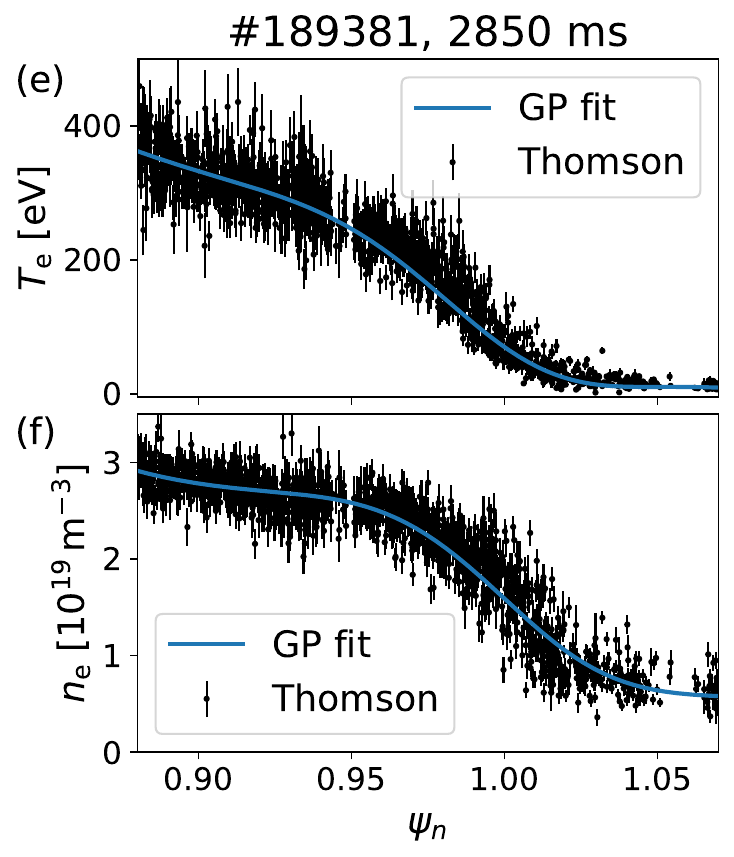}
    \end{subfigure}
    \caption{Electron temperatures and densities in OMP measured by Thomson scattering for (a), (b) H-mode, (c), (d) L-mode, and (e), (f) I-mode discharges. Blue curves indicate the Gaussian process regression fit of each profile.}
    \label{fig:thomson_profiles}
\end{figure*}

\begin{figure*}
    \centering
    \begin{subfigure}{.3\textwidth}
        \centering
        \includegraphics[width=\textwidth]{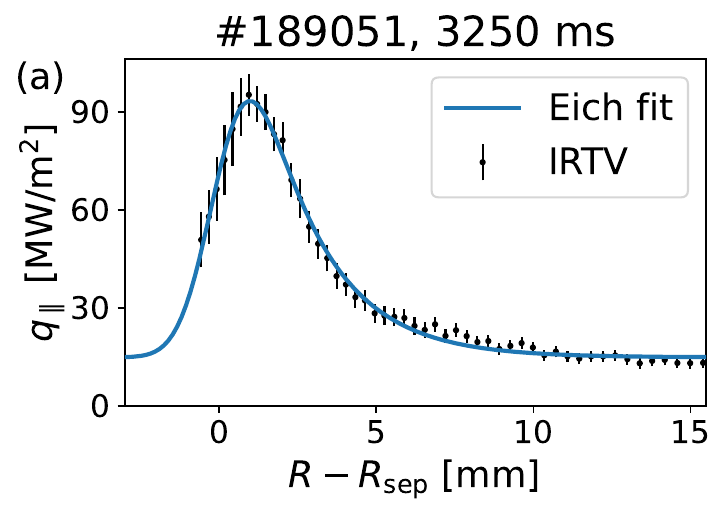}
    \end{subfigure}
    \hfill
    \begin{subfigure}{.3\textwidth}
        \centering
        \includegraphics[width=\textwidth]{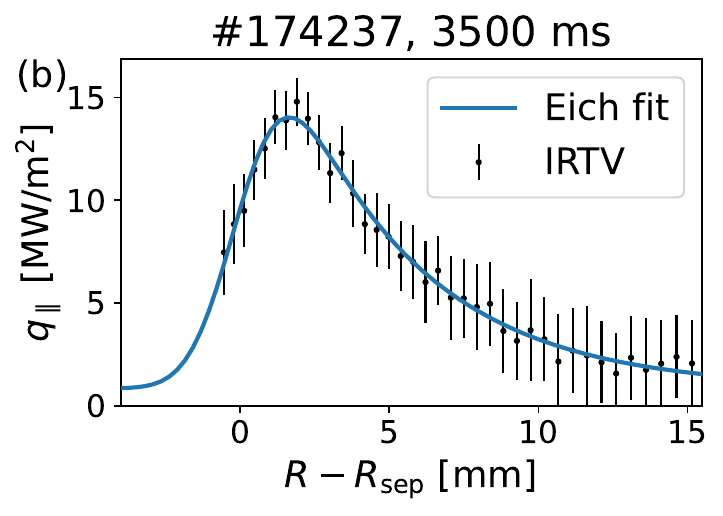}
    \end{subfigure}
    \hfill
    \begin{subfigure}{.3\textwidth}
        \centering
        \includegraphics[width=\textwidth]{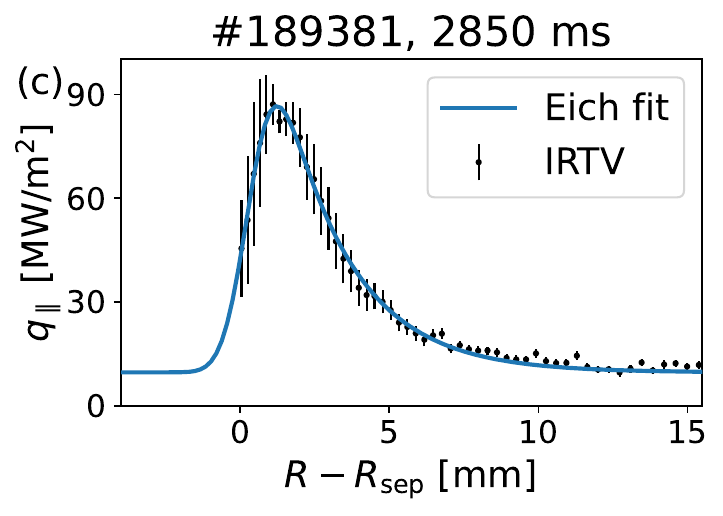}
    \end{subfigure}
    \caption{Parallel heat flux in the outer divertor plate, which is magnetically mapped to the OMP, measured by the IRTV for (a) H-mode, (b) L-mode, and (c) I-mode discharges. Blue curves indicate the Eich fit of each profile.}
    \label{fig:irtv_profiles}
\end{figure*}

Finally, we present the history of the inference process. The trajectories of the loss function, the optimal parameters $(\chi_\mathrm{pf}^*, \chi_\mathrm{sol}^*)$, and their uncertainties in each step during the inference task are shown in Figs.~\ref{fig:bo_history}(a) and (b), respectively. Unlike the gradient-based method, the loss function does not decrease in every step during BO. Nevertheless, the algorithm still learns more information after each step. Noticeably, despite the slow decrease in the loss function, the uncertainty of the parameter $\chi_\mathrm{sol,3}$ increases dramatically in Step 13, suggesting that the GP regression model has detected another possible choice far away from the current optimum. Then in Step 17, the parameter changes rapidly from around $10^{-2}$ to $10^{2}$, accompanied by a substantial decrease in the loss function. The marginal distribution of the parameter $\chi_\mathrm{sol,3}$ is shown in Fig.~\ref{fig:bo_history}(c), which clearly shows the motion of the peak of the distribution function. This transition demonstrates the potential of BO to jump out of a local minimum and serves as a global optimization method.

\section{Tests on experimental data
\label{sec:experimental_data}}

After validation in synthetic data, this section presents the application of our inference method on experimental data from the DIII-D tokamak. The test is performed on three different discharges, including an H-mode (189051 at 3250 ms), an L-mode (174237 at 3500 ms), and an I-mode (189381 at 2850 ms). All three discharges are lower single-null configurations similar to Fig.~\ref{fig:grid_and_diagnostic}, with good diagnostic coverage. The H- and L-mode discharges are in the forward-$B$ configuration (grad-$B$ drift direction towards X point) while the I-mode discharge is in the reverse-$B$ configuration. As discussed above, the inference process uses experimental data from Thomson scattering, IRTV, and bolometer. Here, we first present the details of the data we used in the interpretive model and profiles to be fitted by our inference method, briefly summarize our inference setup, and finally present the inference results. 

\subsection{Experimental data and preprocessing
\label{sec:experimental_data_preprocessing}}

\begin{figure*}
    \centering
    \begin{subfigure}{.32\textwidth}
        \centering
        \includegraphics[width=\textwidth]{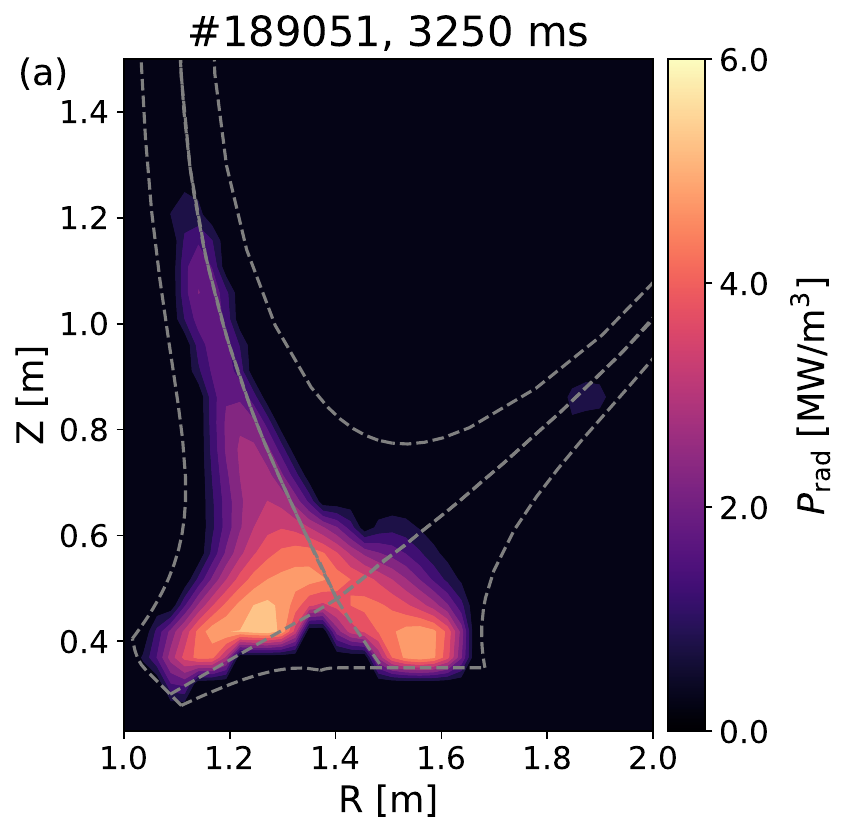}
    \end{subfigure}
    \hfill
    \begin{subfigure}{.32\textwidth}
        \centering
        \includegraphics[width=\textwidth]{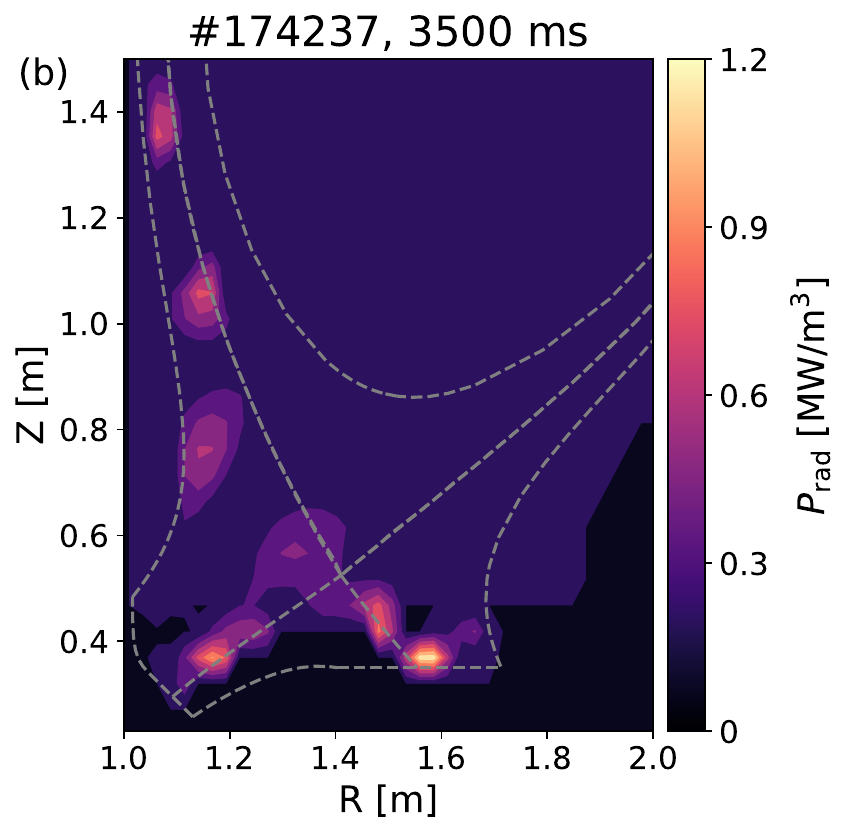}
    \end{subfigure}
    \hfill
    \begin{subfigure}{.32\textwidth}
        \centering
        \includegraphics[width=\textwidth]{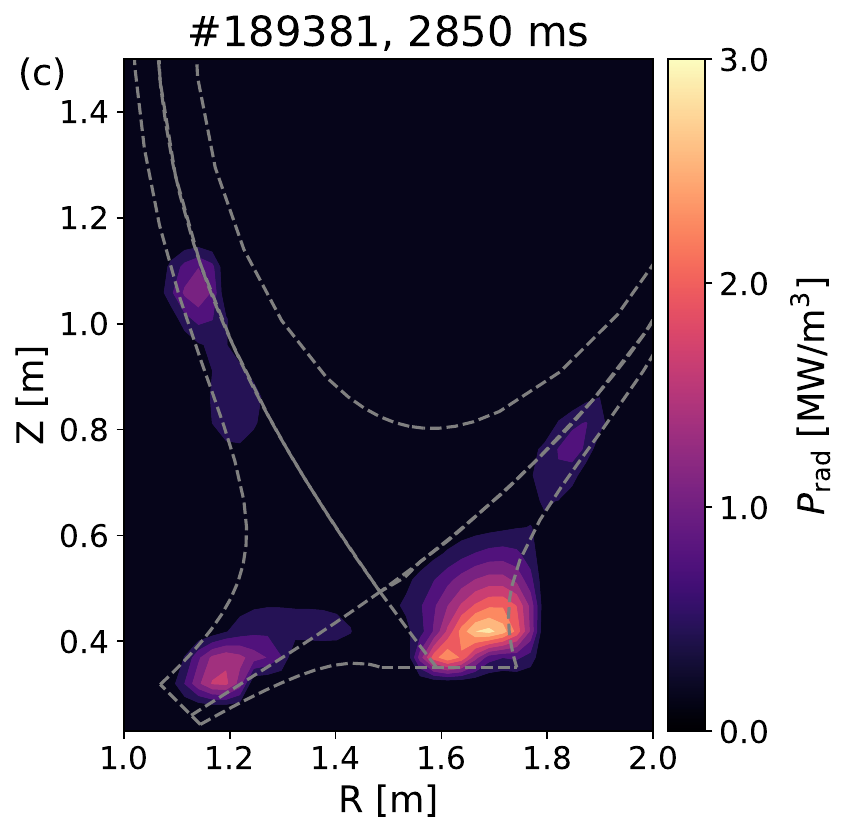}
    \end{subfigure}
    \caption{The radiated power density for (a) H-mode, (b) L-mode, and (c) I-mode discharges obtained from the bolometer. The dashed lines indicate the boundary of UEDGE grids and the separatrix for each discharge.}
    \label{fig:bolometer}
\end{figure*}

Thomson scattering \cite{ponce2010thomson, eldon2012initial} measures both OMP electron temperature and density, which are shown in Fig.~\ref{fig:thomson_profiles}. To get reliable profiles, Thomson data over a short time window is collected. For the H-mode discharge, we collect data within $\pm 200$~ms during 80\% of the ELM cycle. For L- and I-mode discharges, we collect data within $\pm 150$~ms without ELM filtering. New error bars are calculated based on Thomson data to obtain a reasonable probability distribution, which will be discussed in Appendix~\ref{app:error_bar_renormalization} In addition, the temperature and density profiles are fitted using GP regression to obtain continuous $n_\mathrm{e}(\psi)$ and $\mathrm{d}T_\mathrm{e}/\mathrm{d}\psi$, which are used to calculate $\chi_\perp(\psi)$ interpretively within the separatrix using Eq.~(\ref{eq:interpretive_chi}). The density profile at the OMP is also fixed based on the fitted $n_\mathrm{e}(\psi)$ to calculate the parallel momentum balance using Eq.~(\ref{eq:parallel_momentum}).

IRTV \cite{hill1988infrared} measures the plasma-facing components (PFCs) surface temperature. A 2-D heat diffusion equation solver is used to determine the heat flux perpendicular to the PFCs. Then, the heat flux parallel to the magnetic field, shown in Fig.~\ref{fig:irtv_profiles}, is derived from the perpendicular heat flux and magnetically mapped back to the OMP. At each spatial location, the mean and standard deviation are calculated statistically in a short window at the given time slice. For H-mode discharge, data is collected in $\pm 150$~ms during 80\% ELM cycle \footnote{It has to be mentioned that in the H-mode discharge at the specific time slice, the outer strike point (OSP) was moving due to the OSP sweeping. Therefore, we selected the IRTV data in an earlier time window where the location of OSP was stationary and the total amount of heat flux was similar to the current time slice}. For L- and I-mode discharges, the time window is selected as $\pm 100$~ms and $\pm 25$~ms, respectively, without ELM filtering. The parallel heat flux profile can be fitted using the Eich function \cite{eich2011inter,eich2013scaling}, presented in Appendix~\ref{app:eich_fit}. To capture the shape of the heat flux profile, we use the data within five heat flux widths $\lambda_q$. In addition, to count only the conduction heat flux, the fitted background heat flux $q_\mathrm{BG}$, which is assumed to be the radiated power uniformly incident on the surface, is subtracted from the measured value. 

Bolometers \cite{leonard19952d} measure the radiated power along each viewing chord. 2-D radiated power density profiles $P_\mathrm{rad}(R,Z)$, shown in Fig.~\ref{fig:bolometer}, are produced using tomographic reconstruction. The UEDGE grids are chosen to cover most of the radiated power near the plasma edge without intersecting the vessel wall. Although the 2-D radiated power density provides a quantitative description of how radiated power is distributed, it is limited in spatial resolution due to the relatively low number of viewing chords available. So, the radiated power density in some spatial locations might be overestimated or underestimated. As a result, UEDGE may not be able to find an equilibrium of Eq.~(\ref{eq:electron_heat_transport}) if the radiated power in some grid points exceeds the maximum possible heat flux from conduction. To overcome this difficulty, we adopt an \textit{ad hoc} solution to reduce the radiated power in specific grid points by 20\% every time their temperature drops below 1~eV during the UEDGE internal iteration. When a converged equilibrium is found, if only a small fraction ($\lesssim 10\%$) of the radiated power is reduced, we expect this procedure to be valid. If more radiated power needs to be reduced, a more careful investigation is needed to determine the reason behind it. This issue will be discussed later when we present the inference results, especially for the H-mode discharge with the highest radiated power density. 

In addition to the measured profiles, we also calculate the core power $P_\mathrm{core}$ from experimental data as \cite{nelson2021interpretative}
\begin{equation}
    P_\mathrm{core} \doteq 
    P_\mathrm{NBI} + P_\mathrm{Ohmic} - P_\mathrm{ELM} - P_\mathrm{rad,core},
\end{equation}
where $P_\mathrm{NBI}$ is the modulated power from neutral beam injection (NBI), $P_\mathrm{Ohmic}$ is the Ohmic heating power, $P_\mathrm{ELM}$ is the power loss due to the cycle of edge localized modes (ELM) \cite{zohm1996edge}, and $P_\mathrm{rad,core}$ is the radiated power in the core plasma (which is not included in our UEDGE grids).  $P_\mathrm{ELM} \doteq f_\mathrm{ELM}\times E_\mathrm{ELM}$ is calculated as the product of ELM frequency and ELM power, which is absent for ELM-free L- and I-modes. $P_\mathrm{rad,core}$ is obtained by integrating the $P_\mathrm{rad}$ profiles in Fig.~\ref{fig:bolometer} within the UEDGE core boundary. The $P_\mathrm{core}$ for H-, L-, and I-mode discharges used in this paper are 6.07, 0.97, and 2.99~MW.

\subsection{Inference setup}

Following the workflow in Fig.~\ref{fig:BO_workflow}, we briefly summarize the setup of our inference task. We use the interpretive model discussed in Section~\ref{sec:interpretive_model} with radiated power density $P_\mathrm{rad}(R,Z)$ obtained from Fig.~\ref{fig:bolometer}. The core diffusivity $\chi_\perp(\psi<1)$ is calculated interpretively, while the diffusivity in the private flux $\chi_\mathrm{pf}$ and the SOL $\chi_\mathrm{sol}=(\chi_\mathrm{sol,1},...,\chi_\mathrm{sol,4})$ are the five parameters to be inferred. Similar to the synthetic experiment, the inference proceeds after a log transform of the inference variables. 

\begin{figure*}[t]
    \centering
    \begin{subfigure}{.32\textwidth}
        \centering
        \includegraphics[width=\textwidth]{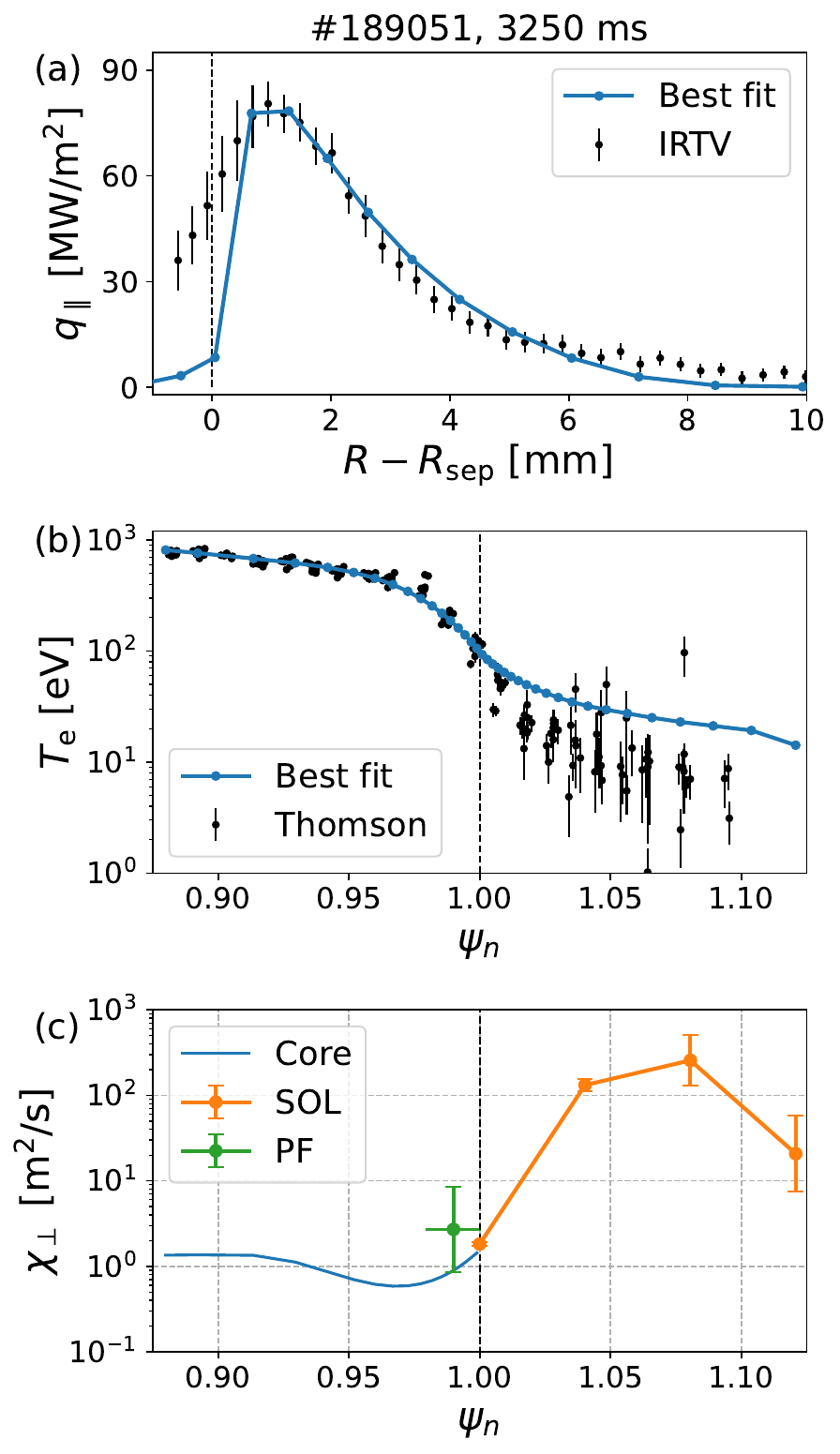}
    \end{subfigure}
    \hfill
    \begin{subfigure}{.313\textwidth}
        \centering
        \includegraphics[width=\textwidth]{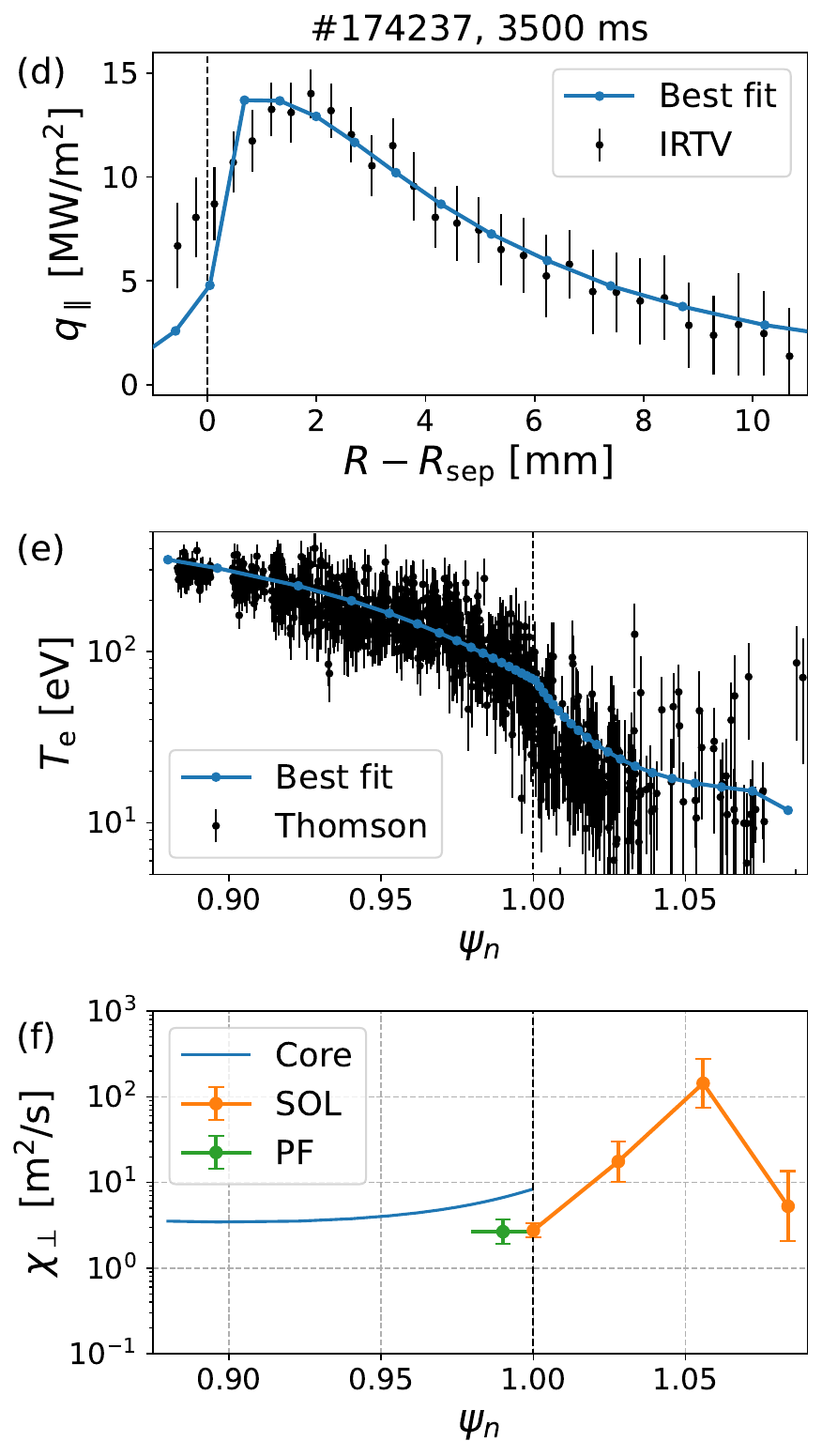}
    \end{subfigure}
    \hfill
    \begin{subfigure}{.313\textwidth}
        \centering
        \includegraphics[width=\textwidth]{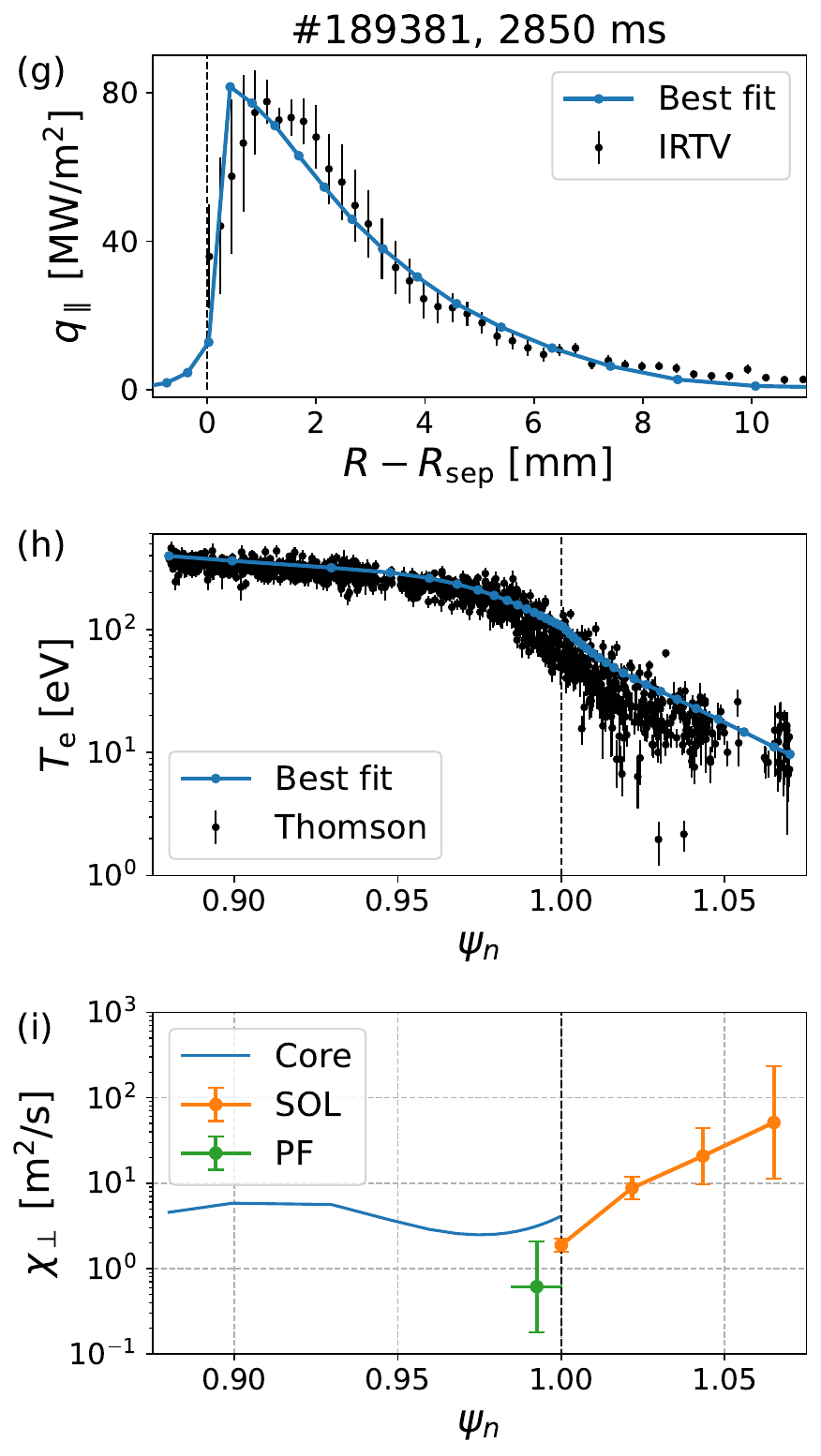}
    \end{subfigure}
    \caption{Inference results for (a-c) H-mode, (d-f) L-mode, and (g-i) I-mode discharges. Figures~(a,d,g) show the parallel heat flux profiles at the outer divertor target, Figs.~(b,e,h) show the electron temperature profiles at the OMP. Figures (c,f,i) show the inferred heat diffusivity. $\chi_\perp$ at plasma core within separatrix is calculated interpretively, while $\chi_\perp$ at SOL and private flux are inferred with uncertainty. }
    \label{fig:bo_result}
\end{figure*}

The electron temperature profile from Thomson scattering and the parallel heat flux profile from IRTV are used to construct the likelihood and loss function. The electron density is not included since it is directly used in calculating the parallel momentum balance from Eq.~(\ref{eq:parallel_momentum}). The likelihood is defined similarly to Eq.~(\ref{eq:chi_square}) and (\ref{eq:likelihood}) but with re-normalized values and error bars discussed in Appendix~\ref{app:error_bar_renormalization}. 

After defining the loss function, we use the same initial sample, run BO process for 30 steps, but calculate 16 samples in parallel in each step. Therefore, UEDGE runs 544 (64 initial samples and $30\times 16$ samples during iteration) different cases during the inference process for each discharge. Due to the inclusion of the radiated power, UEDGE converges more slowly than the synthetic test. For a given heat diffusivity, UEDGE finishes around 10 min for L- and I-modes but needs $15\sim20$~min for H-mode discharges due to its much larger radiated power in the UEDGE grids. So, the inference processes presented in the paper take $\sim 6$ hours for L- and I-modes and $\sim 10$ hours for H-mode. The inference time can be further shortened if more complicated parallelization strategies and larger numbers of computational cores are used.

\subsection{Inference results
\label{sec:inference_results}}

\begin{figure}[t]
    \centering
    \includegraphics[width=0.8\columnwidth]{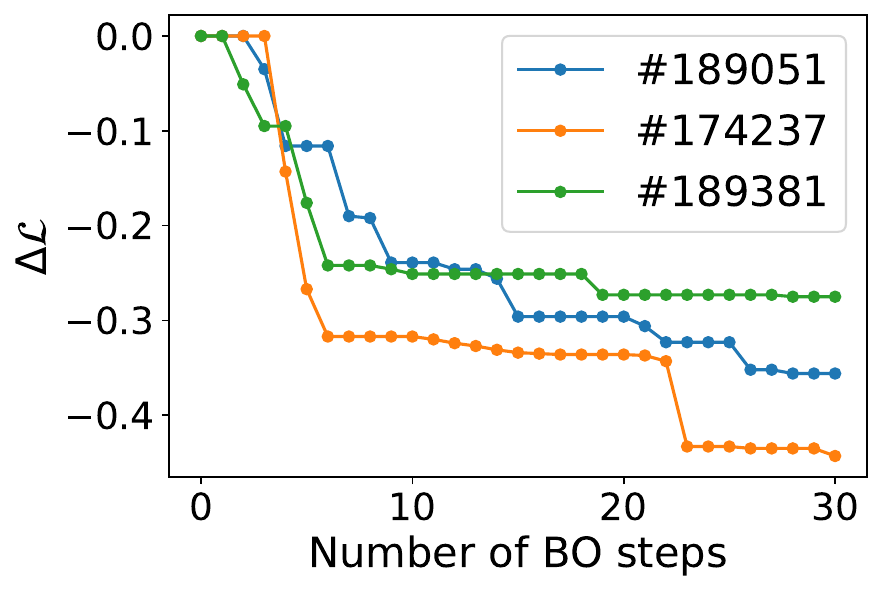}
    \caption{The decrease of the loss functions $\Delta \mathcal{L}$ comparing to the initial losses from random sampling.}
    \label{fig:bo_result_history}
\end{figure}

The decrease of loss functions during the inference for all three discharges is shown in Fig.~\ref{fig:bo_result_history}, where all three inference processes appear to converge after 30 BO steps. The inference results are shown in Fig.~\ref{fig:bo_result}. The heat flux and temperature profiles overlap with the corresponding measurement data, but are not as good as the fitting in the synthetic experiment in Fig.~\ref{fig:synthetic_result}. Noticeably, in the H-mode case, the fitted profile in Fig.~\ref{fig:bo_result}(a) underestimates the heat flux at far SOL, but the profile in Fig.~\ref{fig:bo_result}(b) overestimates the temperature at far SOL. This phenomenon indicates that our simplified interpretive model seems unable to fit both profiles at the same time, while our inference process is trying to find a balance between them. 
In addition, in the L- and I-mode cases, the temperature at separatrix is higher than the average of the Thomson measurement. This might be due to the flux limiter in our interpretive model and will be discussed in Appendix~\ref{app:flux_limiter}. Those mismatches show that the data from experimental measurements are more complicated than synthetic data, where an almost perfect fit always exists. On the other hand, a physics model may not be able to match experimental measurements if the model does not have enough parameters, or may find a lot of equally good matches if the model has too many parameters.

The inferred heat diffusivity $\chi_\perp$ in three discharges is shown at the bottom of Fig.~\ref{fig:bo_result}. $\chi_\perp$ inside the separatrix in three discharges are calculated interpretively and are shown in continuous curves. $\chi_\perp$ in SOL and private flux are inferred with uncertainties. In general, the heat diffusivity just outside the separatrix is best estimated with the smallest uncertainty, similar to the synthetic experiment. The heat diffusivity in the private flux region is estimated with large uncertainty due to the limited measurement data of heat flux in the private flux region. As a result, $\chi_\mathrm{pf}$ barely changes the loss function and cannot be accurately estimated. This is an important feature of BO, which tells the user whether a parameter can be accurately estimated based on the data, whereas other approaches would not.

\begin{figure}[t]
    \centering
    \includegraphics[width=\columnwidth]{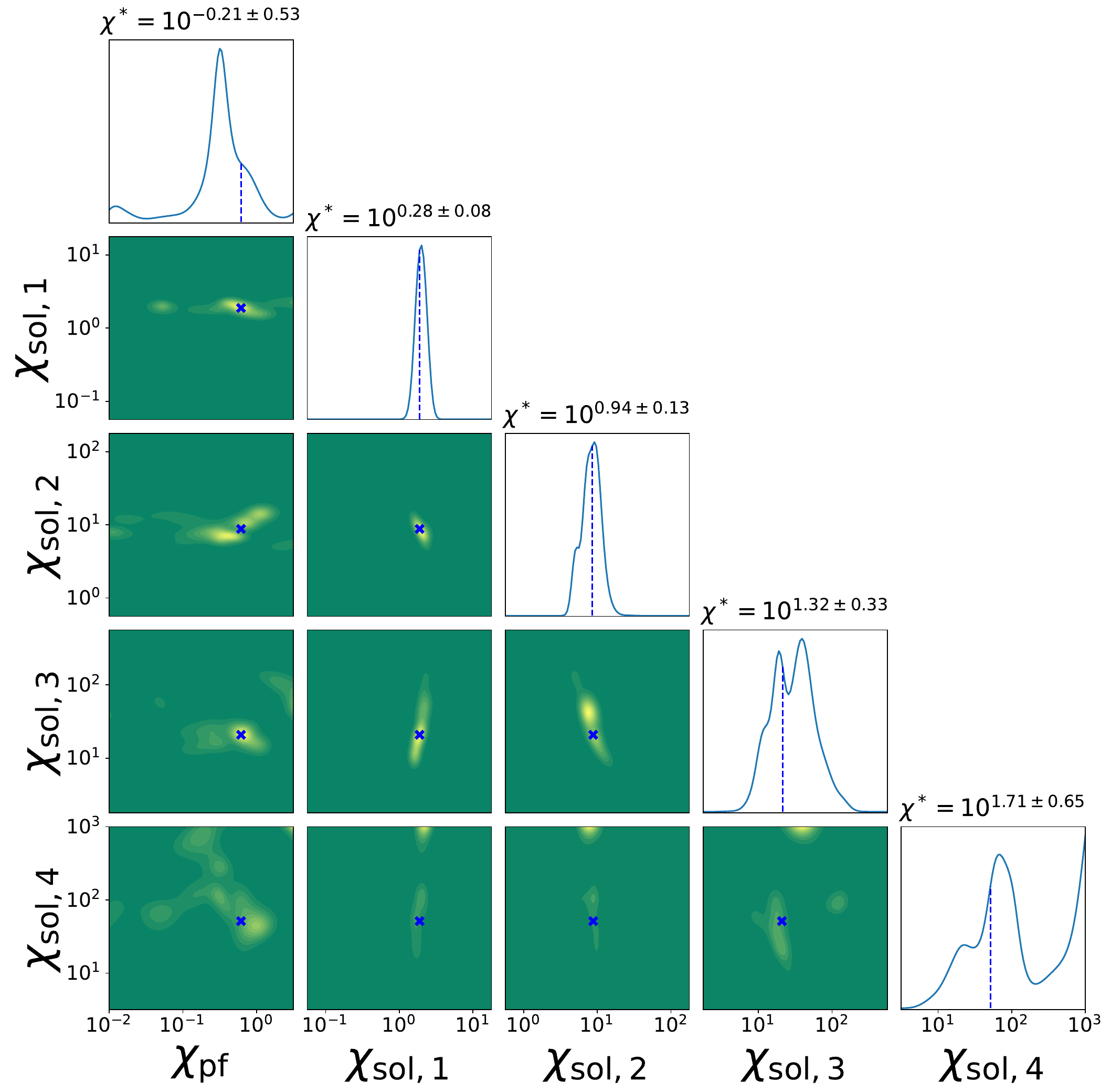}
    \caption{The pair plot of five inference parameters $\chi_\mathrm{pf}, \chi_\mathrm{sol,1}, ..., \chi_\mathrm{sol,4}$ for I-mode discharge. The global optimum parameters are indicated by the dashed lines in the 1-D plots and crosses in the 2-D plots.}
    \label{fig:pair_plot_I_mode}
\end{figure}

We can also visualize the estimated posterior distribution using the pair plot. The distributions of heat diffusivity in the I-mode discharge are shown in Fig.~\ref{fig:pair_plot_I_mode}. Although the distributions in the synthetic experiment in Fig.~\ref{fig:pair_plot} are more Gaussian-like, the distributions obtained from experimental measurements have a more irregular shape and multiple local optima. In particular, the 1-D marginal distribution of the parameter $\chi_\mathrm{sol,4}$ has non-negligible probability within the range $10^1\sim 10^3$, which contributes to its large uncertainty. Another noticeable difference compared to Fig.~\ref{fig:pair_plot} is that the global optimum parameter of the multidimensional distribution function, indicated by the dashed lines in the diagonal plots in Fig.~\ref{fig:pair_plot_I_mode}, may differ from the optimum of the 1-D marginal distributions. These features demonstrate, again, that data from experimental measurements is much more complicated than synthetic data generated from the same physics model.

The estimated $\chi_\perp$ in some regions is relatively large ($>10^2~\mathrm{m^2/s}$) and seems unphysical, which was also observed in other attempts \cite{nelson2021interpretative, canik2011measurements} to estimate the interpretive plasma transport coefficient. This may be due to the assumption that plasma transport is diffusive, while convective transport induced by various drifts \cite{chankin1997classical} is ignored. Intermittent transport due to coherent structures, such as plasma blobs \cite{d2011convective}, can also increase the effective transport coefficient in interpretive models.

\subsection{Radiation and power balance \label{sec:radiation_power_balance}} 

As discussed in Section~\ref{sec:experimental_data_preprocessing}, we have applied an \textit{ad hoc} method to reduce radiated power on some UEDGE grids to ensure convergence. This section analyzes the consequences of such a radiation reduction procedure. 

In the best fits in H-, L-, and I-mode discharges, the total radiated power reductions are 27\%, 6\%, and 11\% compared to measurement, respectively. L- and I-mode discharges have relatively low reduction fractions within expectations, but the H-mode discharge has a relatively high reduction fraction and needs further analysis. The detailed power balances for the H-mode discharge are shown in Tab.~\ref{tab:power_balance_H_mode}, which shows the conduction heat flux at the core, outer boundary, target plate boundaries, and the radiated power. It shows that a large fraction of the power leaves the UEDGE grids through the outer wall, which may be due to the relatively high edge temperature in the fitted profile in Fig.~\ref{fig:bo_result}(b). As a result, there is not enough power left for radiation. 

\begin{table}[ht]
    \centering
    \begin{tabular}{|c | c |} 
        \hline
        Source & Power (MW)  \\[2pt]
        \hline
        Core boundary & 6.08 \\
        Outer boundary & -1.32 \\
        Outer target plate & -0.94 \\
        Inner target plate & -0.13 \\
        Radiation (original) & -4.99 \\ 
        Radiation (reduced) & -3.68 \\
        \hline
    \end{tabular}
    \caption{The power balance in the best-fitted case for H-mode discharge. Positive values indicate power into UEDGE grids, and negative values indicate power out. }
    \label{tab:power_balance_H_mode}
\end{table}

\begin{figure}[t]
    \centering
    \includegraphics[width=0.7\columnwidth]{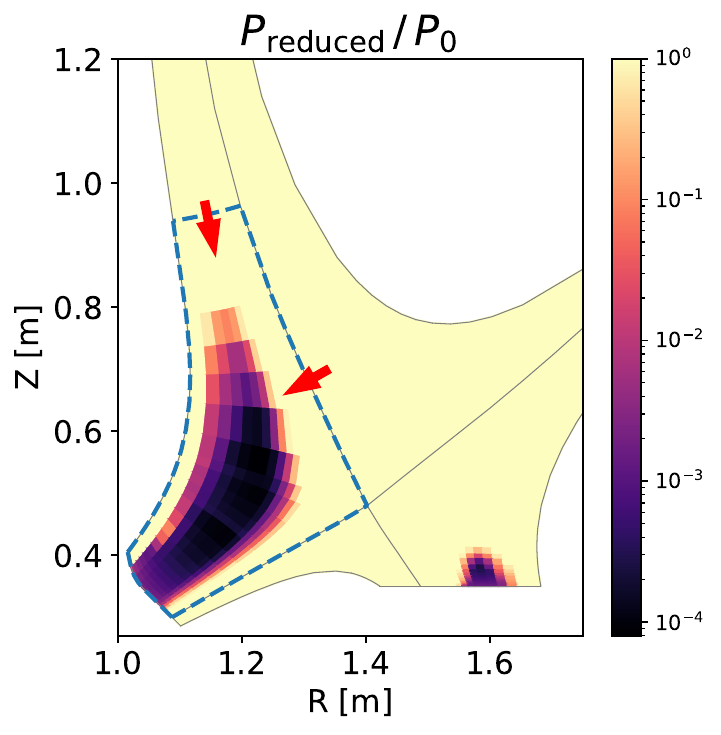}
    \caption{
    The fraction of radiated power used in the model over the measured power in each UEDGE grid. $P_0$ is the original radiated power measured by the bolometer, which is shown in Fig.~\ref{fig:bolometer}, and $P_\mathrm{reduced}$ is the reduced radiated power after the \textit{ad hoc} reduction procedure. The region where radiated power is reduced the most is located in the inner SOL leg, indicated by the blue dashed line. The conduction heat flux into this region is represented by the red arrows.}
    \label{fig:radiation_power_balance}
\end{figure}

To better understand the details of the radiated power reduction, we show the fraction of radiated power reduced in each UEDGE grid in Fig.~\ref{fig:radiation_power_balance}. It shows that most of the radiated power reduced compared to the measurement is localized in the SOL in the inner leg, which coincides with the peak of the radiated power measured in Fig.~\ref{fig:bolometer}. We can further analyze the power balance in the SOL of the inner leg, circled by the dashed blue line. The conduction heat flux into the circled region, represented by the red arrows, is only 1.53~MW. However, the radiated power before reduction is 2.54~MW and much higher than the conduction heat flux. The radiated power after reduction is only 1.38~MW, slightly smaller than the conduction heat flux available. Such a big mismatch cannot be explained by the limited resolution of the bolometer system and indicates a general issue of power balance in our final solution.

There are several factors that may contribute to this power balance issue. From a physical point of view, it is typical for a lower single-null plasma with forward-$B$ direction to have in/out divertor asymmetry, where the inner leg has more radiated power while the outer leg has more heat flux \cite{leonard2018plasma}. This asymmetry is mainly driven by the $E\times B$ drift \cite{chankin1997classical}, which is not included in our interpretive model. This may cause difficulty for our model to balance large radiation with small heat flux in the inner leg. In addition, the temporal resolution of the bolometer system on DIII-D is around tens of milliseconds \cite{leonard19952d}, while the ELM frequency in the discharge is also around 100~Hz. Therefore, bolometer measurement may include some transient events and overestimate the radiated power in a steady state. The uncertainty of the separatrix location \cite{stangeby2015identifying} may affect the separatrix temperature \cite{leonard2017compatibility} and further change the power balance in the SOL. We have perturbed the separatrix location through shifting the midplane profiles within $\pm 4~\mathrm{mm}$ and repeated our inference workflow. However, no major improvement in the power balance was observed. From an optimization point of view, although the profiles in Fig.~\ref{fig:bo_result} are fitted relatively well, the power balance in some discharges is still unsatisfactory. Therefore, we may quantify the ``goodness'' of the power balance and put it into our optimization procedure. For example, require that the fraction of radiated power reduction is less than 20\% and rewrite our optimization task in Eq.~(\ref{eq:optimization}) as a constrained optimization, which can also be solved by the Bayesian optimization method \cite{gammel2024gaussian,gardner2014bayesian}. Detailed exploration of the power balance issue will be left to future work.

\section{Conclusion and discussion
\label{sec:conclusion}}

A workflow of transport coefficient inference has been developed using the Bayesian framework and Bayesian optimization. The Bayesian framework combines experimental observation and uncertainties with the prediction of UEDGE models, which allows us to construct the posterior distribution of the transport coefficients. Parallelized Bayesian optimization serves as an efficient global optimization method to find the most probable transport coefficient. It uses surrogate models to estimate the posterior distribution, which is generally non-Gaussian from experimental data. Uncertainties of the inferred transport coefficients can also be estimated from the posterior distribution. 

To test the inference workflow, we adopt an interpretive model that describes the steady-state electron heat transport in the conduction-limited region dominated by conduction and radiation. Electron heat diffusivity in the SOL and private flux are the inference parameters in this model. The inference workflow is first benchmarked against synthetic data, which successfully recovers the predetermined ground truth within estimation uncertainty. The workflow is then applied to experimental data gathered from H-, L-, and I-mode discharges. With only a few hundred samples, the workflow reproduces the measured temperature and heat flux profiles qualitatively and provides the estimated heat diffusivity with uncertainties. However, experimental profiles are not fitted as good as synthetic profiles because of the complexity of experimental data and the simplicity of the interpretive model. In some cases, the error in the power balance exceeds empirical expectation, highlighting the potential for further refinement of both the physical model and the optimization procedure.

Looking forward, the next step will attempt to apply the inference workflow to more complicated UEDGE models that, for example, include both temperature and density for both electrons and ions. The effect of various drifts should also be investigated in assessing the issue of power balance. However, as the model becomes more complicated, UEDGE may not be able to find an equilibrium solution with given inference parameters, which is a fundamental issue in 2-D transport codes. In those cases, one may need to infer one parameter at a time and fix the other, as discussed in Section~\ref{sec:convergence}. In addition to using a more complicated model, it will also be beneficial to apply the inference workflow to a large database and study the general relationship between the inferred diffusivity and other plasma parameters, such as edge temperature and density.

\begin{acknowledgments}
This work was performed under the auspices of the U.S. Department of Energy by LLNL under Contract DE-AC52-07NA27344, LLNL-JRNL-872895, and received funding from LLNL LDRD 23-ERD-015. This material is also based upon work supported by the U.S. Department of Energy, Office of Science, Office of Fusion Energy Sciences, using the DIII-D National Fusion Facility, a DOE Office of Science user facility, under Award(s) DE-FC02-04ER54698. 
\end{acknowledgments}

\section*{Disclaimer}
This report was prepared as an account of work sponsored by an agency of the United States Government. Neither the United States Government nor any agency thereof, nor any of their employees, makes any warranty, express or implied, or assumes any legal liability or responsibility for the accuracy, completeness, or usefulness of any information, apparatus, product, or process disclosed, or represents that its use would not infringe privately owned rights. Reference herein to any specific commercial product, process, or service by trade name, trademark, manufacturer, or otherwise does not necessarily constitute or imply its endorsement, recommendation, or favoring by the United States Government or any agency thereof. The views and opinions of authors expressed herein do not necessarily state or reflect those of the United States Government or any agency thereof.

\appendix

\section{Error bar renormalization
\label{app:error_bar_renormalization}}

The Thomson scattering data, shown in Fig.~\ref{fig:thomson_profiles}, is collected within a small time window near the given time frame, so some data appear far away from the error bars of each other. In this case, using those data points and error bars directly in Eq.~(\ref{eq:chi_square}) and (\ref{eq:likelihood}) will lead to extremely narrow likelihood functions regardless of model predictions. As a result, the uncertainty we estimate from Eq.~(\ref{eq:uncertainty}) also becomes negligible. To overcome this difficulty, we renormalize the values and error bars of the data from Thomson scattering. 

\begin{figure}[t]
    \centering
    \includegraphics[width=0.8\columnwidth]{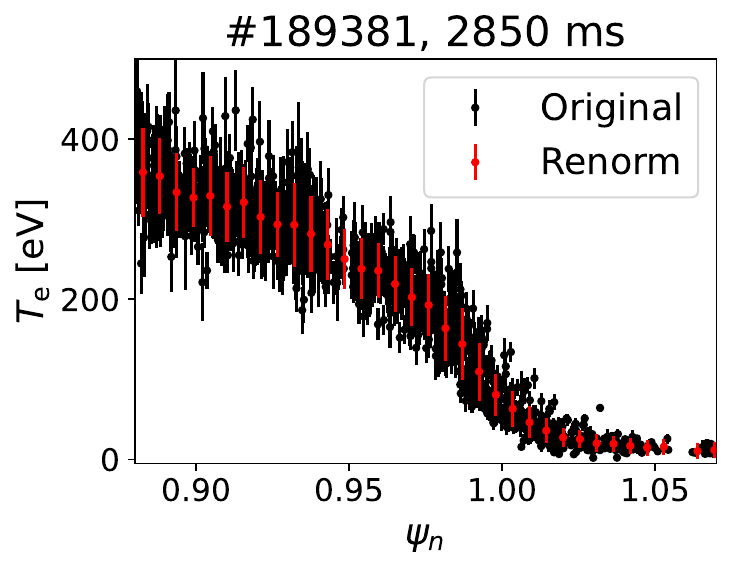}
    \caption{The comparison between original Thomson scattering data and renormalized data for electron temperature at OMP for I-mode discharge.}
    \label{fig:error_renorm}
\end{figure} 

Let the points on the UEDGE grid along the OMP be located at $\psi \in [\psi_a, \psi_b]$. We uniformly separate the domain by points $\{\psi_i\}$, $i=1,...,N_y$, whose numbers equal the UEDGE mesh number. Then, we collect the Thomson scattering data within the range $[\psi_i, \psi_{i+1}]$. Assuming there are $N_i$ data points between each interval, we denote their values and uncertainties as $\{\mu_{ij}, \sigma_{ij}\}$, $j=1,...,N_i$. We consider that each data point represents a Gaussian distribution $\mathcal{N}(\mu_{ij}, \sigma_{ij})$, so the renormalized distribution from $N_i$ data points is 
\begin{equation}
    P_i(x) \doteq \dfrac{1}{N_i}\sum_{j=1}^{N_i}
    \dfrac{1}{\sqrt {2\pi} \sigma_{ij}} \exp\left[{-{\frac {(x-\mu_{ij} )^{2}}{2\sigma_{ij}^{2}}}}\right].
\end{equation}
Then, we can calculate the renormalized mean $\bar{\mu}_i$ and variance $\bar{\sigma}_i$ as
\begin{equation}
        \bar{\mu}_i \doteq \dfrac{1}{N_i} \sum_{j=1}^{N_i} \mu_{ij}, \quad 
        \bar{\sigma}_i^2 \doteq 
        \dfrac{1}{N_i} \sum_{j=1}^{N_i} (\mu_{ij}^2 + \sigma_{ij}^2) - \bar{\mu}_i^2.
\end{equation}
The comparison between the original Thomson data and the renormalized data is shown in Fig.~\ref{fig:error_renorm}, where the renormalized data covers the spreading of the original data.

\section{Eich fit
\label{app:eich_fit}}

The Eich function takes both the exponential radial decay of parallel heat flux and its diffusion to the private flux region, which can be written as
\begin{align}
    q(\bar{s}) = &\frac{q_{0}}{2} \exp
    \left[
        \left(\frac{S}{2\lambda_q}\right)^2 - \frac{\bar{s}}{\lambda_q f_x}
    \right]
    \mathrm{erfc}\left( \frac{S}{2\lambda_q} - \frac{\bar{s}}{S f_x} \right) \nonumber \\
    & + q_\mathrm{BG},
\end{align}
where $\bar{s}=(R_\mathrm{sep} - R) f_x$, $f_x$ is the effective flux expansion factor, $\lambda_q$ is the heat flux width, $S$ is the Gaussian diffusive width, and $q_\mathrm{BG}$ is the background heat flux. The background heat flux $q_\mathrm{BG}$ can be regarded as the value in the limit of $\bar{s}\to\pm\infty$ in Fig.~\ref{fig:irtv_profiles} and will be subtracted from the measured data.

\section{Effect of boundary conditions}

\begin{figure}[b]
    \centering
    \includegraphics[width=0.7\columnwidth]{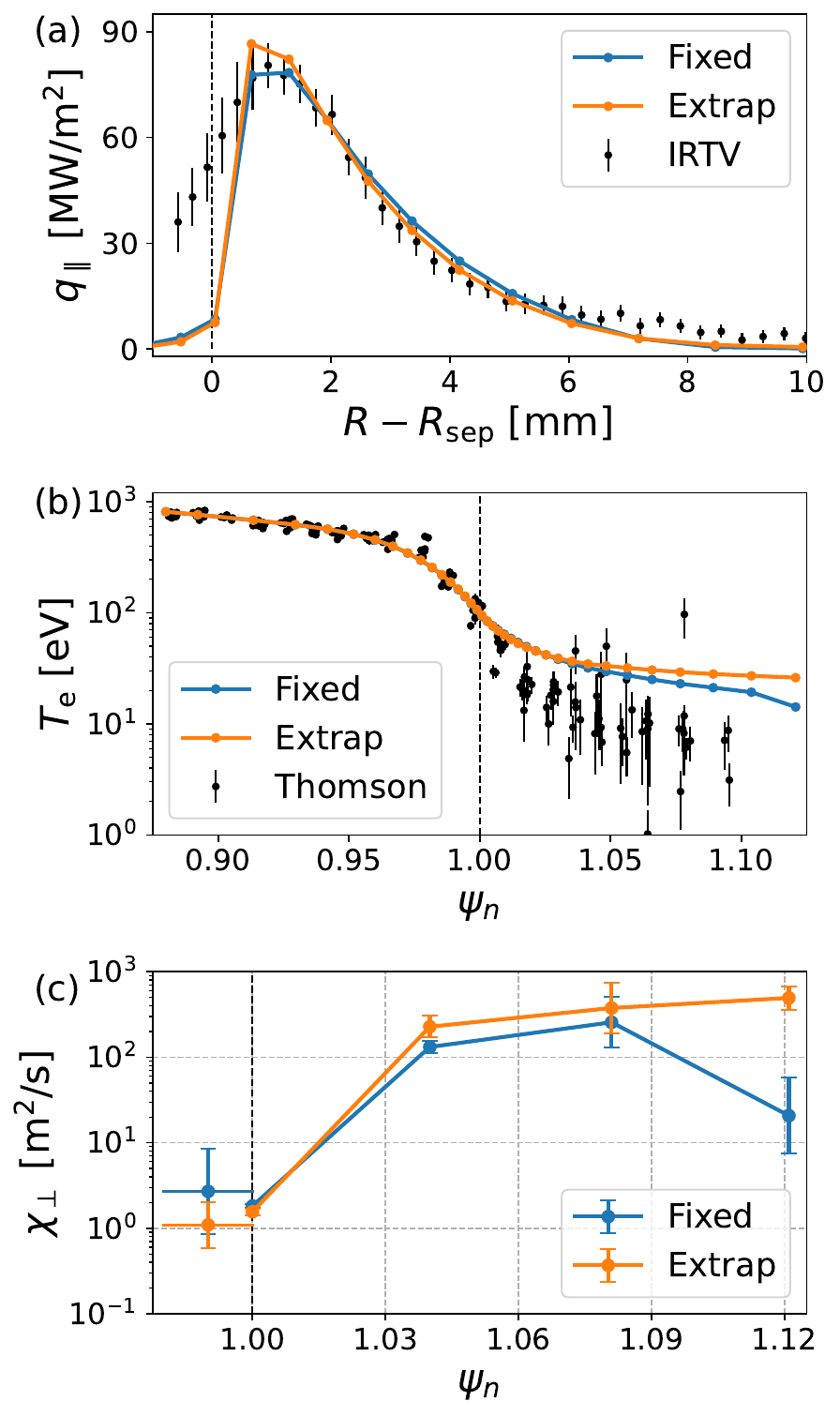}
    \caption{Comparison of fitted profiles and estimated heat diffusivity using two different boundary conditions for H-mode discharge. Blue curves use a fixed temperature boundary condition at the outer wall, orange curves use an extrapolation boundary condition.}
    \label{fig:compare_bc}
\end{figure} 

In the main article, we use the fixed temperature boundary condition for Eq.~(\ref{eq:electron_heat_transport}) at the outer wall boundary. However, the inference result may change when we choose different boundary conditions, especially for the heat diffusivity in the far SOL. Figure~\ref{fig:compare_bc} compares the inference result between the fixed temperature boundary condition and the extrapolation boundary condition for the H-mode discharge. The heat flux profiles from the two boundary conditions are almost identical, whereas the OMP temperature profile differs only in the far SOL, which is an expected difference due to the boundary condition. The inferred heat diffusivity is similar near the separatrix ($\chi_\mathrm{sol,1}$- $\chi_\mathrm{sol,3}$) while it differs in the far SOL ($\chi_\mathrm{sol,4}$) and private flux but shares a similar trend. The results for L- and I-mode discharges share a similar feature. Therefore, the estimated $\chi_\perp$ near the separatrix is more accurate and robust due to its consistency under different boundary conditions.

\section{Effect of flux limiters \label{app:flux_limiter}}

This section discusses the effect of the flux limiter, described by Eq.~(\ref{eq:flux_limiter}), on our fitted profiles. The flux limiter sets the maximum conducting heat flux in the parallel direction based on local density and temperature. In UEDGE, we can turn off the flux limiter by setting the coefficient $\alpha_\mathrm{e}\to\infty$. Like the previous appendix, we rerun our inference process with the flux limiter off and compare it with the fitted profiles in the main article with the flux limiter on. 

\begin{figure}[b]
    \centering
    \begin{subfigure}{.48\columnwidth}
        \centering
        \includegraphics[width=\columnwidth]{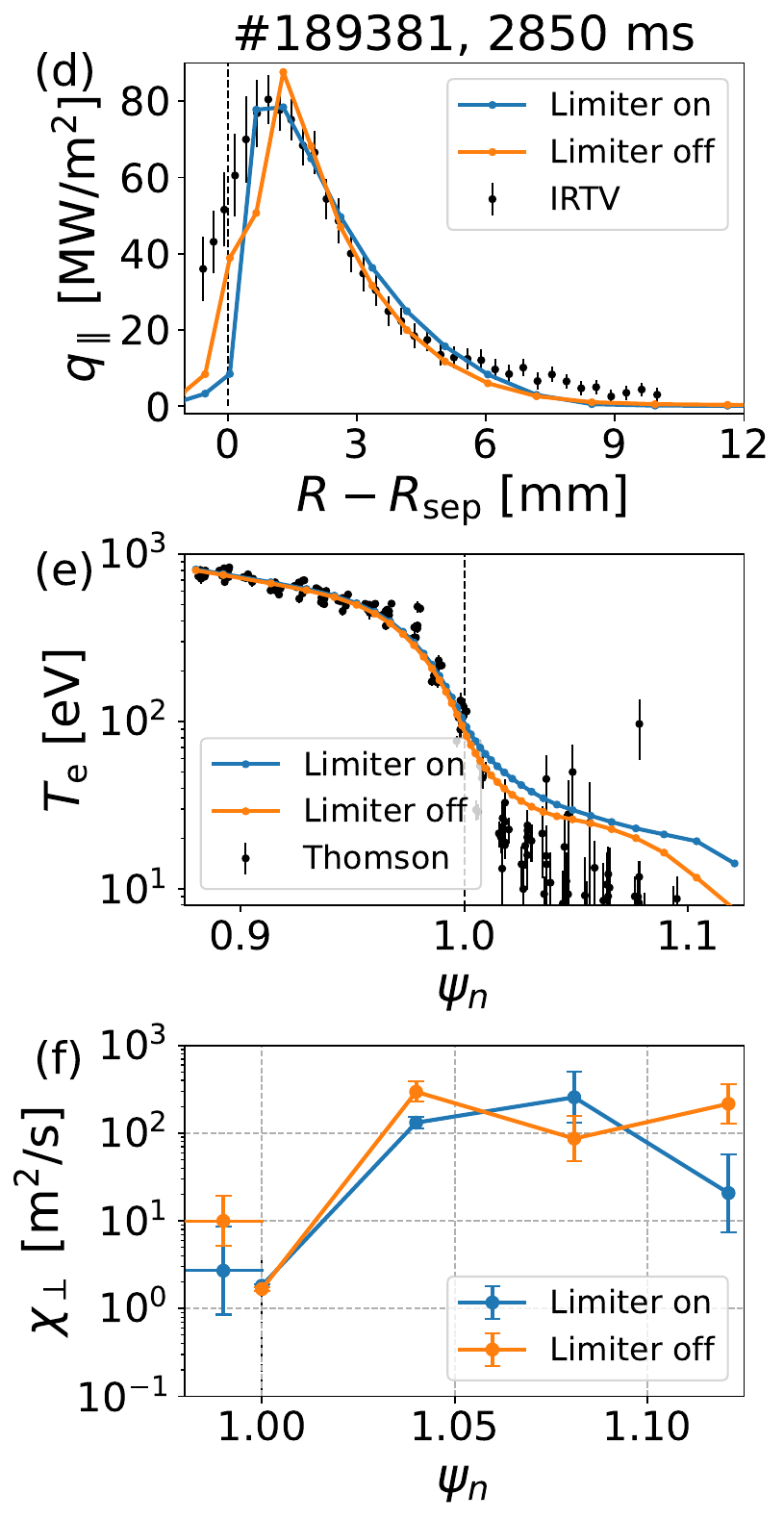}
    \end{subfigure}
    \hfill
    \begin{subfigure}{.48\columnwidth}
        \centering
        \includegraphics[width=\columnwidth]{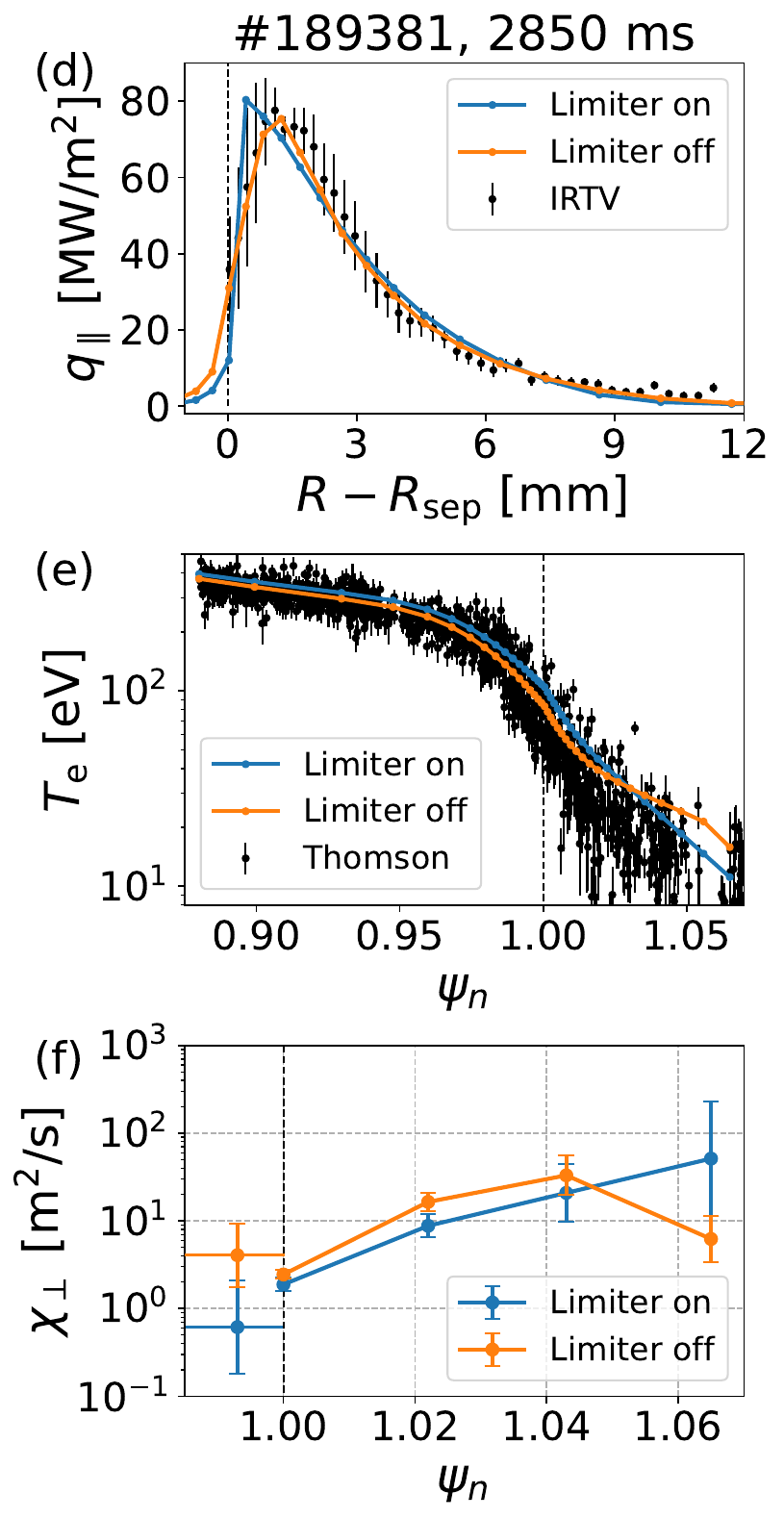}
    \end{subfigure}
    \caption{
    Comparison of fitted profiles and estimated heat diffusivity with (blue) and without (orange) flux limiter for H-mode (189051) and I-mode (189381). The blue curves are identical to Fig.~\ref{fig:bo_result} in the main article.
    }
    \label{fig:flux_limiter}
\end{figure}

\begin{figure}[t]
    \centering
    \begin{subfigure}{.48\columnwidth}
        \centering
        \includegraphics[width=\columnwidth]{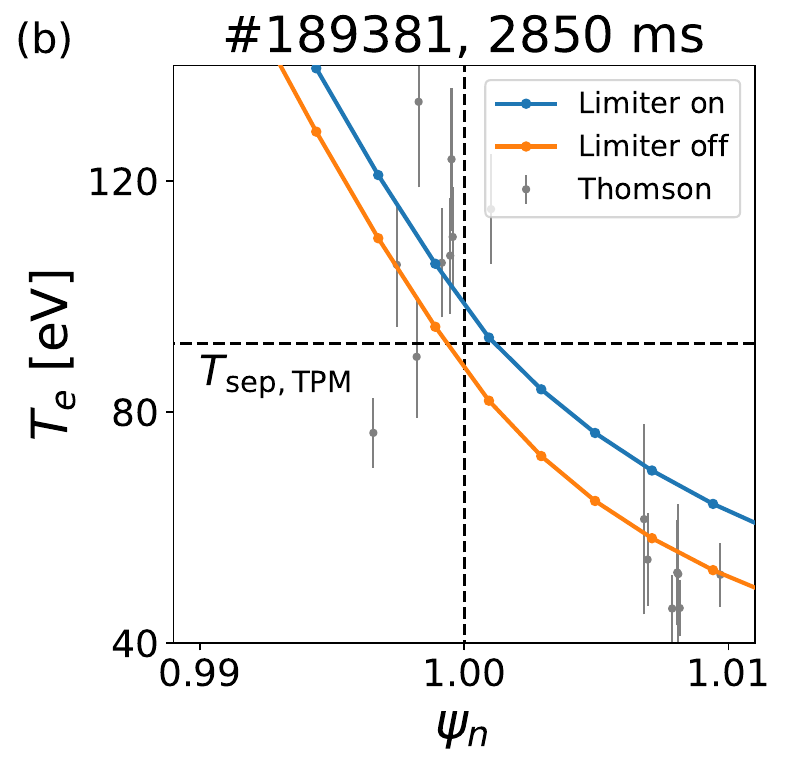}
    \end{subfigure}
    \hfill
    \begin{subfigure}{.48\columnwidth}
        \centering
        \includegraphics[width=\columnwidth]{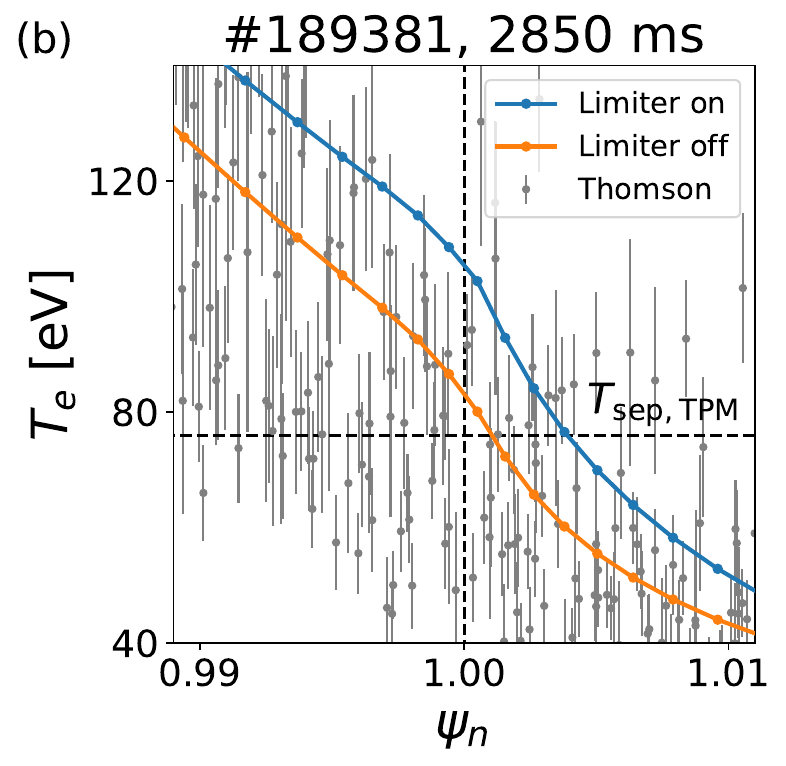}
    \end{subfigure}
    \caption{
    The fitted midplane temperatures near the separatrix with and without the flux limiter.
    }
    \label{fig:tsep}
\end{figure}

The comparison of inference results in H-mode and I-mode discharges is shown below in Fig.~\ref{fig:flux_limiter}. In the H-mode case, both parallel heat flux and midplane temperature profiles are fitted with similar quality regardless of the flux limiter. In the I-mode case, although the parallel heat flux profiles appear similar, the midplane temperature profile with flux limiter is higher than the profile without flux limiter by $\sim15$~eV from the core boundary to near SOL. A closer comparison of midplane temperature near the separatrix is shown in Fig.~\ref{fig:tsep}. The separatrix temperatures, $T_\mathrm{sep}$, calculated from a simplified two-point model (TPM) \cite{leonard2017compatibility}, are also shown in the figure. It is clear that in the H-mode case, the fitted separatrix temperature is closer to the TPM prediction. In the I-mode case, however, the fitted separatrix temperature with the flux limiter is significantly higher. 

The difference between the inferred profiles with and without a flux limiter can be understood as follows. When the flux limiter is turned on, the heat flux will be reduced. As a result, the separatrix temperature has to be increased to maintain the level of heat flux to match the IRTV measurement. Although H- and I-mode discharges have similar amounts of parallel heat flux in Fig.~\ref{fig:flux_limiter} (a) and (d), H-mode discharge has much higher separatrix density than I-mode in Fig.~\ref{fig:thomson_profiles} (b) and (d). Since $q_\mathrm{fl}$ in Eq.~(\ref{eq:limited_heat_flux}) is proportional to plasma density, the heat flux limiting effect is more significant in I-mode than H-mode. As discussed in Ref.~\onlinecite{leonard2017compatibility}, the flux limiter has a very small effect on the TPM calculation. Therefore, the inferred separatrix temperature without a flux limiter is closer to the TPM prediction. This indicates that kinetic effects may play a role in fitting the profiles of low-density discharges, which needs further exploration in future studies.

\bibliography{main.bbl}
\bibliographystyle{unsrt.bst}

\end{document}